\documentclass[reprint,
superscriptaddress,
amsmath,amssymb,
aps, 
pra,
twocolumn
]{revtex4-2}

\usepackage{graphicx}
\usepackage{fancyhdr}
\usepackage{xcolor}
\usepackage{hyperref}
\usepackage{amsmath}
\usepackage{amsthm}
\usepackage{amssymb}
\usepackage[normalem]{ulem}

\usepackage[dvipsnames]{xcolor}

\usepackage{orcidlink}

\begin{document}

\title{Generalized Rotating-Wave Approximations Hierarchy for the Interacting Two-Qubit Rabi Model}
\author{Ricardo Herrera Romero\,\orcidlink{0009-0008-7596-6148}}
\affiliation{Departamento de F\'isica, Universidad Aut\'onoma Metropolitana-Iztapalapa, Av. Ferrocarril San Rafael Atlixco 186, C.P. 09310 Mexico City, Mexico}
\author{Carlos A. Gonz\'alez-Guti\'errez\,\orcidlink{0000-0002-1734-1405}}
\affiliation{Instituto de Ciencias Físicas, Universidad Nacional Autónoma de México, Av. Universidad s/n, Col. Chamilpa, Cuernavaca, Morelos, 62210 México}
\author{M. A.~Bastarrachea-Magnani\,\orcidlink{0000-0002-1552-4101}}
\affiliation{Departamento de F\'isica, Universidad Aut\'onoma Metropolitana-Iztapalapa, Av. Ferrocarril San Rafael Atlixco 186, C.P. 09310 Mexico City, Mexico}
\email{bastarrachea@xanum.uam.mx}

% ============================================================
% Abstract
% ============================================================

\begin{abstract}
We investigate the two-qubit quantum Rabi model for identical qubits symmetrically coupled to a single bosonic mode and interacting with each other through transverse and longitudinal couplings. We extend the adiabatic, generalized rotating-wave (GRWA), and squeezed generalized rotating-wave (SGRWA) approximations to include the qubit-qubit interactions, including self-consistent and variational choices of the bosonic parameters. The resulting analytical hierarchy allows us to
distinguish the interplay between both kinds of qubit-qubit interactions and the effects of displacement, neighboring-manifold mixing, and squeezing on the
low-energy properties. We find that the transverse interaction mainly changes the collective-spin structure of the states, while the longitudinal interaction also modifies the displacement-dependent oscillator overlaps and introduces an additional two-boson coupling. We compare the analytical results with numerically exact diagonalization for the low-energy spectrum, ground-state energy, mean photon number, two-qubit concurrence, and spin correlations. The results provide a compact analytical description of how direct qubit--qubit interactions modify both the spectrum and the eigenstate structure beyond the rotating-wave approximation for few-qubit light-matter systems.
\end{abstract}


\maketitle

% ============================================================
% Introduction
% ============================================================

\section{Introduction}

The quantum Rabi model provides one of the simplest descriptions of light--matter interaction, modeling a two-level system coupled to a single bosonic mode~\cite{Rabi1936}. Beyond its original role in quantum optics, the model has become increasingly relevant in modern quantum-technology platforms where light--matter coupling can reach a sizable fraction of the characteristic bare frequencies, particularly in the ultra-strong coupling (USC) regime~\cite{Larson2021,FornDiaz2019,Kockum2019}. Closely related qubit--oscillator descriptions arise in cavity and circuit quantum electrodynamics, molecular systems, and spin--boson settings~\cite{Stepanov2008,Liu2009,Gardas2011,Herfort2001,Eisfeld2005,Ashhab2010,FornDiaz2010}, with superconducting circuits providing a prominent platform for accessing strong light—matter coupling~\cite{Wallraff2004,Niemczyk2010,Fedorov2010,Yoshihara2017,Blais2021}. 

Near resonance and at weak coupling, the rotating-wave approximation (RWA) neglects counter-rotating processes and reduces the Rabi model to the analytically solvable Jaynes--Cummings model~\cite{Jaynes1963,Shore1993}. However, this description becomes insufficient in the ultrastrong-coupling (USC) regime, where the light--matter coupling is no longer negligible compared with the characteristic system frequencies and counter-rotating terms modify both the spectrum and eigenstates~\cite{Larson2021,FornDiaz2019,Kockum2019}. The excitation number is then no longer conserved, giving rise to effects such as Bloch--Siegert shifts, virtual bosonic dressing, and nontrivial ground-state correlations~\cite{FornDiaz2019,Kockum2019}. Despite substantial progress toward exact descriptions of the single- and two-qubit Rabi models~\cite{Braak2011,duan2015,Peng2014,Wang2014}, compact analytical approximations remain valuable for tractable, physically transparent descriptions beyond the RWA.

Displaced-oscillator methods provide a systematic set of analytical approaches to capture the phenomenology of the USC regime. In the adiabatic approximation, the Hamiltonian is expressed in a spin-dependent displaced basis, retaining only couplings within the same oscillator manifold~\cite{irish2005,irish2007}. The Generalized Rotating-Wave Approximation (GRWA) extends this approximation by restoring the dominant transitions between neighboring displaced manifolds, improving the description over a broader range of coupling strengths and detunings~\cite{irish2007,zhang2015}. The squeezed GRWA (SGRWA) further supplements the displacement with a squeezing transformation, allowing the oscillator basis to adjust both its center and its fluctuations~\cite{zhangyu2016,gonzalez2021}. These schemes can also be combined with self-consistent or variational bosonic parameters~\cite{Albert2011,Mao2019,Zhang2011-1,yu2012}. Together, they provide a natural hierarchy for separating the roles of displacement, inter-manifold hybridization, and squeezing.

The two-qubit Rabi model extends this setting to the multi-emitter case, allowing field-mediated correlations to emerge beyond the RWA. Its spectral and entanglement properties have been studied with perturbative, Bargmann-space, extended coherent-state, and GRWA-based methods ~\cite{Chilingaryan2013,Peng2012,Peng2014,duan2015,zhang2015,Lee2013}. Related two-qubit Rabi systems have also been considered in the context of nonequilibrium dynamics, dynamical quantum phase transitions, and effective spin--spin interactions~\cite{DiBello2024,Grimaudo2023}. In superconducting architectures, the qubits can also interact directly through capacitive or inductive elements, tunable couplers, or auxiliary resonator modes~\cite{Blais2003,Majer2007,Niskanen2007,Allman2010,Blais2021}, giving rise to transverse, longitudinal, or more general spin--spin couplings~\cite{Richer2017,Royer2017}. These additional interactions can reshape the spectral and correlation properties of the system~\cite{Grimaudo2023,JiaoPu2026}. However, most GRWA-based treatments neglect direct qubit--qubit couplings, leaving open how they modify the displaced and squeezed structure underlying the analytical approximations themselves.

In this work, we extend this analytical hierarchy to two identical, symmetrically coupled qubits with direct transverse and longitudinal interactions. Exchange symmetry reduces the problem to the symmetric spin-one sector, within which we formulate conventional, self-consistent, and variational versions of the adiabatic, GRWA, and SGRWA schemes. We benchmark these approximations against numerically exact diagonalization using the low-energy spectrum, mean photon number, concurrence, and spin correlations, thereby probing both eigenvalues and wave-function structure. We find that the transverse interaction acts mainly through the collective-spin structure, whereas the longitudinal interaction also modifies displacement-dependent oscillator matrix elements and generates an additional two-boson coupling. The comparison further shows that displacement optimization, inter-manifold mixing, and squeezing provide complementary corrections, while the direct interactions modify the parameter ranges over which each approximation remains accurate.

The paper is organized as follows. In Sec.~\ref{sec:2} we introduce the two-qubit Rabi model with qubit-qubit interactions, and its reduction to the symmetric spin-one sector. In Sec.~\ref{sec:3} we revisit the theoretical framework behind the approximations discussed in this work: the adiabatic approximation, the GRWA, and the SGRWA  together with their variational extensions. In Sec.~\ref{sec:4} we define the observables used to characterize the approximate eigenstates, and in Sec.~\ref{sec:5} we compare the analytical hierarchy with numerically exact diagonalization. Sec.\ref{circuit::qed} is dedicated to the discussion of the corresponding circuit-QED implementation of the model, and the experimentally accesible values of the parameters involved. Finally, in Sec.~\ref{sec:6} we summarize the main results, provide further discussion and concluding remarks. Additional derivations are provided in several Appendices.

% ============================================================
% Model
% ============================================================

\section{The Interacting Two-Qubit Rabi Model}\label{sec:2} 

We consider two identical qubits symmetrically coupled to a single bosonic mode and qubit-qubit couplings. Setting $\hbar=1$, the Hamiltonian reads 

\begin{gather} 
    \hat{H}=\omega \hat{a}^{\dagger}\hat{a}+\frac{\omega_0}{2}\left(\hat{\sigma}_{z}^{1}+\hat{\sigma}_{z}^{2}\right)+\frac{g}{2}\left(\hat{\sigma}_{x}^{1}+\hat{\sigma}_{x}^{2}\right)(\hat{a}^{\dagger}+\hat{a})+\nonumber \\ 
    \frac{\eta_x}{2}\hat{\sigma}_{x}^{1}\hat{\sigma}_{x}^{2}+\frac{\eta_z}{2}\hat{\sigma}_{z}^{1}\hat{\sigma}_{z}^{2}. \label{eq:interacting_two_qubit_qrm}
\end{gather} 

Here, $\omega$ and $\omega_0$ denote the oscillator and qubit frequencies, respectively, while $\hat{a}$ and $\hat{a}^{\dagger}$ are the bosonic annihilation and creation operators, satisfying $[\hat{a},\hat{a}^{\dagger}]=1$. The operators $\hat{\sigma}_{\mu}^{i}$, with $\mu=x,y,z$, are Pauli operators acting on qubit $i$. The parameter $g$ is the collective light--matter coupling, whereas $\eta_x$ and $\eta_z$ characterize the direct transverse and longitudinal qubit--qubit interactions.

Direct qubit--qubit couplings are naturally generated in superconducting-circuit architectures. They may arise from capacitive or inductive elements, tunable couplers, or auxiliary resonator modes~\cite{Blais2003,Majer2007,Niskanen2007,Hime2006,Allman2010,Blais2021}. Depending on the circuit design and working basis, these mechanisms can produce transverse exchange-type interactions, longitudinal interactions, or more general effective spin--spin couplings~\cite{Richer2017,Royer2017,Blais2021}. The parameters $\eta_x$ and $\eta_z$ provide an effective description of direct qubit--qubit interactions that will compete with the dressing induced by the bosonic mode.

Since the qubits are identical and symmetrically coupled, it is convenient to introduce the collective operators

\begin{gather} 
    \hat{J}_{\mu}=\frac{1}{2}\left(\hat{\sigma}_{\mu}^{1}+\hat{\sigma}_{\mu}^{2}\right),\qquad \mu=x,y,z, \label{eq:collective_operators} 
\end{gather}

which give 

\begin{gather} 
    \hat{H}=\omega \hat{a}^{\dagger}\hat{a}+\omega_0\hat{J}_{z}+g\hat{J}_{x}(\hat{a}^{\dagger}+\hat{a})\nonumber \\  +\eta_x\hat{J}_{x}^{2}+\eta_z\hat{J}_{z}^{2}+E_S^{0}\hat{\mathbb{I}}, \label{eq:collective_spin_hamiltonian}
\end{gather} 
with $E_S^{0}=-(\eta_x+\eta_z)/2$. Exchange symmetry separates the two-qubit Hilbert space into an antisymmetric singlet ($j=0$) and a totally symmetric triplet ($j=1$) sector. In the singlet subspace, the collective-spin operators vanish, $\hat J_{\mu}|0,0\rangle=0$ for $\mu=x,y,z$, so this state does not couple to the oscillator. Its eigenstates are therefore $|0,0\rangle\otimes|n\rangle$, with energies $E_{S,n}=n\omega+E_S^{0}$. It should be noted that although the singlet does not hybridize with the triplet sector, its energy may cross the symmetric-sector energy levels as the qubit interactions are varied, so whenever the global ground state of the complete system is discussed, the competence between the lowest triplet and the singlet energies must be considered. 

The analytical treatment can then be restricted to the symmetric $j=1$ sector, where the two qubits behave as an effective spin-one degree of freedom. Within this  sector, the system can
be interpreted as a two-emitter Dicke-type model supplemented by quadratic collective-spin
interactions~\cite{Dicke1954,Tavis1968,Garraway2011,zhang2015}. 

The constant $E_S^{0}$ only shifts the entire spectrum, so we omit it in the calculations below and restore it only when absolute energies of the original Hamiltonian are required. The working Hamiltonian is then
\begin{gather}
    \hat{H}_{j=1}=\omega \hat{a}^{\dagger}\hat{a}+\omega_0\hat{J}_{z}+g\hat{J}_{x}(\hat{a}^{\dagger}+\hat{a})+\eta_x\hat{J}_{x}^{2}+\eta_z\hat{J}_{z}^{2}.
\label{eq:working_hamiltonian}
\end{gather}

Within this sector, the transverse interaction $\eta_x\hat{J}_x^2$ acts along the same collective direction as the light--matter coupling and mixes spin components, whereas the longitudinal term $\eta_z\hat{J}_z^2$ is diagonal in the $\hat{J}_z$ basis and shifts the relative energies of the $m=0$ and $m=\pm1$ states. Their competition with the oscillator-induced displacement determines the low-energy structure analyzed below.

% ============================================================
% Approximations
% ============================================================

\section{Analytical Approximations}
\label{sec:3}

We now construct a hierarchy of analytical approximations for the symmetric low-energy sector of
the interacting two-qubit Rabi model. Our starting point is the displaced-oscillator approach
underlying the adiabatic approximation and the generalized rotating-wave approximation
(GRWA)~\cite{irish2005,irish2007,zhang2015,zhangyu2016}, together with the squeezed extension introduced~\cite{zhangyu2016}.  We adapt these
approximations to the present model with direct transverse and longitudinal qubit–qubit
interactions, parametrized by $\eta_{x}$ and $\eta_{z}$, respectively.

From a bird's-eye view, at the lowest level, the adiabatic approximation retains only processes within a fixed displaced-oscillator manifold. Then, the GRWA subsequently restores the dominant transitions between neighboring manifolds, while the SGRWA widens the bosonic basis thanks to the deformation in quantum fluctuations. For each approximation, we consider two cases: fixed parameters and a variational prescription optimized from the ground-state energy. This organization separates two conceptually different improvements. Variational optimization modifies the bosonic basis in which the Hamiltonian is represented, whereas the progression from the adiabatic approximation to the GRWA and SGRWA changes the physical processes retained after the transformation. This distinction will be important when comparing ground-state and excited-state properties below. 

Hence, we first derive the resulting transformed Hamiltonian and then specify the truncation and bosonic-parameter prescription associated with each approximation. We introduce the so-called polaron transformation~\cite{Alcalde2012,Jaako2016,DeBernardis2018,Pilar2020}
\begin{gather}
    \hat U_{\beta} = \exp\left[ \beta \hat J_x(\hat a^\dagger-\hat a) \right],
    \label{eq:Ubeta}
\end{gather}
which gives
\begin{gather}
    \hat H(\beta) = \omega\hat a^\dagger\hat a + (g-\omega\beta)\hat J_x(\hat a^\dagger+\hat a) + K(\beta)\hat J_x^2  + \nonumber\\ \label{eq:Hbeta}
    \omega_0 \left[ \hat J_z \cosh\!\left[\beta(\hat a^\dagger-\hat a)\right] - i\hat J_y \sinh\!\left[\beta(\hat a^\dagger-\hat a)\right] \right] \\
    + \eta_z \left[ \hat J_z \cosh\!\left[\beta(\hat a^\dagger-\hat a)\right] - i\hat J_y \sinh\!\left[\beta(\hat a^\dagger-\hat a)\right] \right]^2 \nonumber,
\end{gather}
where
\begin{gather}
    K(\beta)=\omega\beta^2-2g\beta+\eta_x .
\label{eq:Kbeta}
\end{gather}

%The transformed Hamiltonian makes the interaction asymmetry explicit: $\eta_x$ enters through the collective-spin sector, whereas $\eta_z$ also generates displacement-dependent oscillator matrix elements. The specific terms retained from Eq.~\eqref{eq:Hbeta} determine the approximations developed below.

We observe that the transverse interaction, $\eta_x$, acts directly within the
collective-spin sector through the $J_x^2$ term and modifies the corresponding
spin-sector energies and eigenstates. By contrast, the longitudinal interaction, $\eta_z$, generates additional displacement-dependent bosonic matrix elements after the transformation. These two interactions therefore affect the analytical approach in qualitatively different ways and constitute the main extensions of the standard two-qubit GRWA considered below.

% ============================================================
% Adiabatic
% ============================================================

\subsection{Adiabatic approximation}
\label{subsec:adiabatic}

The term ``adiabatic'' refers here to the separation of energy scales between
the oscillator and the qubit degrees of freedom. In the regime where the
oscillator dynamics is fast compared with the characteristic qubit dynamics,
the bosonic state can approximately adjust to a given collective-spin
configuration. This motivates the use of a spin-dependent displaced-oscillator
basis~\cite{irish2005,irish2007,zhang2015}.

Within the adiabatic approximation, matrix elements of the transformed Hamiltonian connecting different oscillator occupations are neglected \cite{irish2005,irish2007,zhang2015}. The problem therefore separates into independent displaced manifolds labeled by the quantum number of the oscillator $n$. The diagonal oscillator overlaps required for the interacting model are
\begin{gather}
    G_n(\beta) = e^{-\beta^2/2}L_n(\beta^2), \qquad A_n(\beta) = e^{-2\beta^2}L_n(4\beta^2),
\label{eq:adiabatic_factors}
\end{gather}
where $L_n(x)$ denotes a Laguerre polynomial. Projecting Eq.~\eqref{eq:Hbeta} onto the $n$th manifold gives
\begin{gather}
    \hat H^{\mathrm{Ad}}_n(\beta)= \omega n + K(\beta)\hat J_x^2 + \omega_0G_n(\beta)\hat J_z + \nonumber\\  \frac{\eta_z}{2}A_n(\beta) \left(\hat J_z^2-\hat J_y^2\right) + \frac{\eta_z}{2} \left(\hat J_z^2+\hat J_y^2\right).
\label{eq:Hadiabatic}
\end{gather}
The corresponding three energy branches are
\begin{gather}
E^{\mathrm{Ad}}_{n,0}(\beta) = \omega n + K(\beta) + \frac{\eta_z}{2} \left[1-A_n(\beta)\right], \label{eq:EadA} \\
E^{\mathrm{Ad}}_{n,\pm}(\beta) = \omega n + \frac{K(\beta)}{2} + \frac{3\eta_z}{4} + \frac{\eta_z}{4}A_n(\beta)  
\nonumber\\ 
\pm \frac{1}{2} \sqrt{ \left[ K(\beta) + \frac{\eta_z}{2} \left(A_n(\beta)-1\right) \right]^2 + 4\omega_0^2G_n^2(\beta)
}.
\label{eq:Eadpm}
\end{gather}
Equations~\eqref{eq:EadA} and~\eqref{eq:Eadpm} define the adiabatic spectrum for an arbitrary displacement $\beta$. The fixed, self-consistent, and variational prescriptions used to determine the bosonic parameters are discussed together below in Sec.~\ref{subsec:parameter_prescriptions}.

% ============================================================
% GRWA
% ============================================================

\subsection{Generalized rotating-wave approximation}
\label{subsec:grwa}

The adiabatic approximation treats each displaced-oscillator manifold independently and neglects transitions between different oscillator sectors. The GRWA improves this description by retaining the dominant one-boson transitions between neighboring manifolds in the displaced basis~\cite{irish2007,zhang2015}.

For an arbitrary oscillator manifold $n$, it is convenient to define
\begin{gather}
    h_n=\omega_0G_n, \\  
    d_n=\frac{K}{2}+\frac{\eta_z}{4}(A_n+3), \\  q_n=\frac{K}{2}+\frac{\eta_z}{4}(A_n-1), \\  e_n=K+\frac{\eta_z}{2}(1-A_n), \label{eq:hdq_n} 
\end{gather}
where the diagonal overlaps $G_n=G_n(\beta)$ and $A_n=A_n(\beta)$ were introduced in Eq.~\eqref{eq:adiabatic_factors}. These quantities contain the spin contribution associated with each displaced-oscillator manifold.
To construct a common collective-spin basis for the GRWA blocks, we use the lowest oscillator manifold, $n=0$, as the reference sector. Its spin Hamiltonian in the $\hat J_z$ representation is
\begin{gather}
    \hat H_s^{(0)} = \begin{pmatrix}
    d_0+h_0 & 0 & q_0\\
    0 & e_0 & 0\\
    q_0 & 0 & d_0-h_0
    \end{pmatrix}.
\label{eq:Hs0GRWA}
\end{gather}
Defining
\begin{gather}
    R=\sqrt{h_0^2+q_0^2},
    \quad
    \epsilon_{\pm}=d_0\pm R,
    \quad
    \epsilon_0=e_0,
    \nonumber\\
    \cos\theta=\frac{h_0}{R},
    \quad
    \sin\theta=\frac{q_0}{R},
\label{eq:thetaGRWA}
\end{gather}
The quantities $\epsilon_-$, $\epsilon_0$, and $\epsilon_+$ are the eigenenergies of the reference spin Hamiltonian. We refer to the corresponding states as dressed collective-spin states because they diagonalize the spin sector after the oscillator displacement, rather than the bare $\hat J_z$ Hamiltonian. We notice that these energies are not, in general, the eigenvalues of the full GRWA blocks once inter-manifold couplings are retained.
The corresponding eigenstates of the reference spin Hamiltonian are
\begin{gather}
    |-\rangle =
    -\sin\frac{\theta}{2}|1\rangle + \cos\frac{\theta}{2}|-1\rangle,
    \\ 
    |+\rangle = \cos\frac{\theta}{2}|1\rangle + \sin\frac{\theta}{2}|-1\rangle.
\label{eq:rotatedSpinStates}
\end{gather}
Together with $|0\rangle$, these states define the collective-spin basis used to organize all oscillator manifolds. Projecting the spin Hamiltonian of an arbitrary manifold $n$ onto this dressed basis gives the diagonal contributions
\begin{gather}
    \xi_{-,n} = d_n-h_n\cos\theta-q_n\sin\theta, \qquad
    \xi_{0,n}=e_n, \\ \nonumber
    \xi_{+,n} = d_n+h_n\cos\theta+q_n\sin\theta.
\label{eq:xiGRWA}
\end{gather}
The essential GRWA correction is the retention of the leading matrix elements connecting neighboring oscillator manifolds. The relevant one-boson overlaps are
\begin{gather}
    R_n(\beta) = \frac{\beta}{\sqrt{n+1}} e^{-\beta^2/2} L_n^{(1)}(\beta^2), \\ 
    B_n(\beta) = \frac{2\beta}{\sqrt{n+1}} e^{-2\beta^2} L_n^{(1)}(4\beta^2),
\label{eq:RBn}
\end{gather}
which give

\begin{gather}
    Z_n = \frac{1}{\sqrt{2}} \left[ (g-\omega\beta)\sqrt{n+1}(c-s)  + \omega_0R_n(c+s) \right. \nonumber \\
    \left.- \frac{\eta_z}{2}B_n(c-s) \right],
    \label{eq:Zn}
    \\
    Y_n = \frac{1}{\sqrt{2}} \left[ (g-\omega\beta)\sqrt{n+1}(c+s)  + \omega_0R_n(c-s) \right. \nonumber \\
    \left. + \frac{\eta_z}{2}B_n(c+s) \right],
\label{eq:Yn}
\end{gather}
where $c=\cos(\theta/2)$ and $s=\sin(\theta/2)$. In addition, the longitudinal interaction generates a two-boson matrix element,
\begin{gather}
W_n = \frac{\eta_z}{4}A_n^{(2)}\cos\theta, \\  A_n^{(2)} =
 \frac{4\beta^2}{\sqrt{n(n+1)}} e^{-2\beta^2} L_{n-1}^{(2)}(4\beta^2),
\label{eq:Wn}
\end{gather}
which is absent for $\eta_z=0$ and constitutes an additional coupling generated by the interacting extension.
For $n\geq1$, the retained GRWA states are organized into the three-dimensional manifold
\begin{gather}
    \mathcal{B}_n = \left\{ |-,n+1\rangle,\, |0,n\rangle,\, |+,n-1\rangle \right\},
\label{eq:GRWAbasis}
\end{gather}
with effective Hamiltonian
\begin{gather}
    \hat H_n^{\mathrm{GRWA}} =
    \begin{pmatrix}
    V_{-,n} & Z_n & W_n\\
    Z_n & V_{0,n} & Y_{n-1}\\
    W_n & Y_{n-1} & V_{+,n}
\end{pmatrix},
\label{eq:HGRWA}
\end{gather}
where
\begin{gather}
    V_{-,n} = \omega(n+1)+\xi_{-,n+1}, \\ 
    V_{0,n} = \omega n+\xi_{0,n}, \\ 
    V_{+,n} = \omega(n-1)+\xi_{+,n-1}.
\label{eq:VGRWA}
\end{gather}
The three GRWA energies in each retained manifold, $E_{n,j}^{\mathrm{GRWA}}$ with $j=1,2,3$, are the eigenvalues of Eq.~\eqref{eq:HGRWA} and therefore satisfy
\begin{gather}
    \det\!\left[ E_{n,j}^{\mathrm{GRWA}}\mathbb I - \hat H_n^{\mathrm{GRWA}} \right] = 0.
\label{eq:GRWAsecular}
\end{gather}
The explicit solution of the resulting cubic equation is given in App.~\ref{app:grwa_matrix_elements}.
The lower boundary of the oscillator ladder requires separate treatment because states with negative oscillator occupation do not exist. The state $|-,0\rangle$ therefore remains decoupled from the reduced block and has energy $\epsilon_-$, while the remaining states form
\begin{gather}
    \mathcal{B}_0 = \left\{ |0,0\rangle,\, |-,1\rangle \right\}, \\ 
    \hat H_0^{\mathrm{GRWA}} =
    \begin{pmatrix}
    \epsilon_0 & Z_0\\
    Z_0 & \omega+\xi_{-,1}
\end{pmatrix}.
\label{eq:HGRWA0}
\end{gather}
Its two eigenvalues are
\begin{gather}    E_{0,\pm}^{\mathrm{GRWA}} = \frac{ \epsilon_0+\omega+\xi_{-,1} }{2} \nonumber\\ 
\pm \frac{1}{2} \sqrt{ \left( \epsilon_0-\omega-\xi_{-,1} \right)^2 + 4Z_0^2 }.
\label{eq:EGRWA0}
\end{gather}
The lower GRWA spectrum is therefore determined by the branch $\epsilon_-$ and the eigenvalues $E_{0,\pm}^{\mathrm{GRWA}}$, together with the eigenvalues of the higher retained blocks. Their role in the ground-state selection depends on the prescription adopted for $\beta$ and is discussed in Sec.~\ref{subsec:parameter_prescriptions}.

% ============================================================
% SGRWA
% ============================================================

\subsection{Squeezed generalized rotating-wave approximation}
\label{subsec:sgrwa}

The SGRWA extends the GRWA by allowing the oscillator basis to adjust both its displacement and its width through an additional bosonic squeezing transformation~\cite{zhangyu2016},
\begin{gather}
    \hat V_{\lambda} = \exp\!\left[ \lambda(\hat a^2-\hat a^{\dagger 2}) \right], \qquad \hat W_{\beta,\lambda} = \hat V_{\lambda}\hat U_{\beta}.
\label{eq:squeezeTransformation}
\end{gather}
The GRWA truncation scheme is otherwise unchanged: the same diagonal, neighboring-manifold, and longitudinally induced two-boson matrix elements are retained, while the explicit $\hat a^2$ and $\hat a^{\dagger 2}$ terms generated by the transformed oscillator Hamiltonian are neglected.
The effect of squeezing can be incorporated through
\begin{gather}
    \bar{\beta}=\beta e^{-2\lambda}, \; E_{\lambda}=\omega\sinh^2(2\lambda), \; \Omega_{\lambda}=\omega\cosh(4\lambda),
\label{eq:squeezedParameters}
\end{gather}
together with the replacement
\begin{gather}
    \left\{
    G_n^\lambda,
    A_n^\lambda,
    R_n^\lambda,
    B_n^\lambda,
    A_n^{(2),\lambda}
    \right\} =
    \left.
    \left\{
    G_n,
    A_n,
    R_n,
    B_n,
    A_n^{(2)}
    \right\}
    \right|_{\beta\rightarrow\bar{\beta}} .
\label{eq:squeezedFactors}
\end{gather}
Accordingly, the diagonal quantities $h_n^\lambda$, $d_n^\lambda$, $q_n^\lambda$, and $e_n^\lambda$ follow directly from Eq.~\eqref{eq:hdq_n} after replacing $G_n\rightarrow G_n^\lambda$ and $A_n\rightarrow A_n^\lambda$. As in the GRWA, the lowest oscillator manifold defines the reference collective-spin basis,
\begin{gather}
    R_\lambda = \sqrt{(h_0^\lambda)^2+(q_0^\lambda)^2}, \; \cos\theta_\lambda = \frac{h_0^\lambda}{R_\lambda}, \; \sin\theta_\lambda = \frac{q_0^\lambda}{R_\lambda},
\label{eq:thetaSGRWA}
\end{gather}
and the diagonal spin contributions are obtained from Eq.~\eqref{eq:xiGRWA} as
\begin{gather}
    \xi^\lambda_{-,n} = d_n^\lambda - h_n^\lambda\cos\theta_\lambda - q_n^\lambda\sin\theta_\lambda, \\ \xi^\lambda_{0,n}
      =e_n^\lambda, \\ \xi^\lambda_{+,n} = d_n^\lambda + h_n^\lambda\cos\theta_\lambda + q_n^\lambda\sin\theta_\lambda .
\label{eq:xiSGRWA}
\end{gather}
Writing $c_\lambda=\cos(\theta_\lambda/2)$ and $s_\lambda=\sin(\theta_\lambda/2)$, and defining
\begin{gather}
    \Gamma_n^\lambda = (g-\omega\beta)e^{2\lambda}\sqrt{n+1},
\label{eq:GammaSGRWA}
\end{gather}
the neighboring-manifold couplings become
\begin{gather}
    Z_n^\lambda = \frac{1}{\sqrt{2}} \left[ \Gamma_n^\lambda(c_\lambda-s_\lambda) + \omega_0R_n^\lambda(c_\lambda+s_\lambda) \right. \nonumber \\
    \left. - \frac{\eta_z}{2}B_n^\lambda(c_\lambda-s_\lambda) \right],
    \label{eq:Zlambda}
    \\
    Y_n^\lambda = \frac{1}{\sqrt{2}} \left[ \Gamma_n^\lambda(c_\lambda+s_\lambda) + \omega_0R_n^\lambda(c_\lambda-s_\lambda)\right. \nonumber \\
    \left. + \frac{\eta_z}{2}B_n^\lambda(c_\lambda+s_\lambda) \right],
\label{eq:Ylambda}
\end{gather}
while the longitudinally induced two-boson coupling retains the GRWA form,
\begin{gather}
    W_n^\lambda = \frac{\eta_z}{4} A_n^{(2),\lambda}\cos\theta_\lambda .
\label{eq:Wlambda}
\end{gather}
For $n\geq1$, the SGRWA preserves the manifold structure of Eq.~\eqref{eq:GRWAbasis}, with $|\pm\rangle$ replaced by the corresponding squeezed reference-spin eigenstates. The effective Hamiltonian therefore reads
\begin{gather}
\hat H_n^{\mathrm{SGRWA}}
=
\begin{pmatrix}
    V_{-,n}^{\lambda} & Z_n^\lambda & W_n^\lambda\\
    Z_n^\lambda & V_{0,n}^{\lambda} & Y_{n-1}^\lambda\\
    W_n^\lambda & Y_{n-1}^\lambda & V_{+,n}^{\lambda}
\end{pmatrix},
\label{eq:HSGRWA}
\end{gather}
where
\begin{gather}
    V_{-,n}^{\lambda} = E_\lambda + \Omega_\lambda(n+1) + \xi^\lambda_{-,n+1}, \\ V_{0,n}^{\lambda} = E_\lambda + \Omega_\lambda n + \xi^\lambda_{0,n}, \\ V_{+,n}^{\lambda} = E_\lambda + \Omega_\lambda(n-1) + \xi^\lambda_{+,n-1}.
\label{eq:VSGRWA}
\end{gather}
The three SGRWA energies $E_{n,j}^{\mathrm{SGRWA}}$, $j=1,2,3$,
are obtained from
\begin{gather}
    \det\!\left[ E_{n,j}^{\mathrm{SGRWA}}\mathbb I - \hat H_n^{\mathrm{SGRWA}} \right] = 0.
\label{eq:SGRWAsecular}
\end{gather}
At the lower boundary, the same reduction as in the GRWA applies.
The state $|-,0\rangle_\lambda$ remains decoupled from the reduced block, with energy $E_\lambda+d_0^\lambda-R_\lambda$, while
\begin{gather}
    \hat H_0^{\mathrm{SGRWA}} =
    \begin{pmatrix}
    E_\lambda+e_0^\lambda
    &
    Z_0^\lambda\\
    Z_0^\lambda
    &
    E_\lambda+\Omega_\lambda+\xi_{-,1}^\lambda
\end{pmatrix}.
\label{eq:HSGRWA0}
\end{gather}
Its eigenvalues are
\begin{gather}
    E_{0,\pm}^{\mathrm{SGRWA}} = E_\lambda + \frac{ e_0^\lambda+\Omega_\lambda+\xi_{-,1}^\lambda }{2} 
    \nonumber \\
    \pm \frac{1}{2} \sqrt{ \left( e_0^\lambda-\Omega_\lambda-\xi_{-,1}^\lambda \right)^2 + 4(Z_0^\lambda)^2 }.
\label{eq:ESGRWA0}
\end{gather}

% ============================================================
% Parameters
% ============================================================

At this stage, including the squeezing transformation yields the same hierarchy of bosonic-parameter prescriptions introduced for the displaced GRWA. The essential idea remains the one underlying the original GRWA, i.e., the accuracy of the subsequent rotating-wave-like truncation depends strongly on the choice of bosonic basis~\cite{irish2007}.
However, the introduction of the two independent deformations, the displacement
$\beta$ and a squeezing parameter $\lambda$ improves substantially the descriptions towards the USC regime~\cite{zhangyu2016}.

\subsection{Fixed, self-consistent, and variational parameters}
\label{subsec:parameter_prescriptions}

The approximations above specify the retained matrix elements, while the bosonic parameters determine the basis for the corresponding truncation. Here, we discuss the three prescriptions for determining the parameters: fixed, self-consistent, and fully variational.

\textit{Fixed prescription.}
For the displaced adiabatic and GRWA schemes, the conventional polaron choice is
\begin{gather}
\beta_p=\frac{g}{\omega}.
\label{eq:fixed_beta}
\end{gather}
This provides the reference displaced basis against which the optimized prescriptions are compared.

\textit{Self-consistent prescriptions.}
Following the displaced GRWA approach ~\cite{irish2007} and its squeezed extension ~\cite{zhangyu2016}, we determine the bosonic parameters self-consistently for the present interacting model. This is a useful intermediate improvement that can be employed without performing the full variational
optimization. These schemes retain the displaced representation underlying the GRWA while allowing the qubit energy scale to oppose the oscillator displacement~\cite{Zhang2011-1,yu2012}. For the displaced schemes, this yields
\begin{gather}
    \beta_{\rm sc} = \frac{g}{ \omega + \omega_0e^{-\beta_{\rm sc}^2/2}
      -\eta_z e^{-2\beta_{\rm sc}^2} }.
\label{eq:beta_sc}
\end{gather}
The correction to the noninteracting Rabi problem appears in the denominator and originates from the qubit-induced overlap between displaced oscillator states. $\eta_{z}$ renormalizes the same effective scale that controls the oscillator displacement. The transverse interaction $\eta_{x}$ does not appear explicitly in this equation because, for the displaced-state energy from which the condition is obtained, it contributes only through a $\beta$-independent term. When squeezing is included, the displacement and squeezing parameters are determined simultaneously from
\begin{gather}
    \beta_{\rm sc} = \frac{g}{ \omega + \omega_0e^{-4\lambda_{\rm sc}} e^{-\frac{\beta_{\rm sc}^2}{2}e^{-4\lambda_{\rm sc}}} - \eta_z e^{-4\lambda_{\rm sc}} e^{-2\beta_{\rm sc}^2e^{-4\lambda_{\rm sc}}} },
    \label{eq:sgrwa_sc_beta}
    \\
    \lambda_{\rm sc} = \frac{1}{8} \ln\left[
    1+ \frac{2\beta_{\rm sc}^2}{\omega} \left( \omega_0 e^{ -\frac{\beta_{\rm sc}^2}{2}e^{-4\lambda_{\rm sc}}} - \eta_z e^{-2\beta_{\rm sc}^2e^{-4\lambda_{\rm sc}}} \right) \right].
    \label{eq:sgrwa_sc_lambda}
\end{gather}
In the weak-displacement and weak-squeezing limit, these coupled equations give
\begin{gather}
    \beta_{\rm S} \simeq \frac{g}{\omega+\omega_0-\eta_z}, \qquad \lambda_{\rm S} \simeq \frac{\omega_0-\eta_z}{4\omega} \left( \frac{g}{\omega+\omega_0-\eta_z} \right)^2 .
\label{eq:approx_squeezed_parameters}
\end{gather}
Thus, the self-consistent prescriptions adapt the displaced and squeezed bosonic basis to the interacting problem without modifying the GRWA and SGRWA block structures.

\textit{Variational prescriptions.}
The fully variational schemes instead minimize the retained physical energy branches. For the adiabatic approximation,
\begin{gather}
    E_{\rm GS}^{\rm VAd} = \min_{\beta} \left\{ E^{\rm Ad}_{0,-}(\beta), E^{\rm Ad}_{0,0}(\beta) \right\}.
\label{eq:VAd_prescription}
\end{gather}
For the GRWA and SGRWA,
\begin{gather}
    E_{\rm GS}^{\rm VGRWA} = \min_{\nu\in\mathcal R_{\rm GRWA}} \left[ \min_{\beta} E_\nu^{\rm GRWA}(\beta) \right], \\ E_{\rm GS}^{\rm VSGRWA} = \min_{\nu\in\mathcal R_{\rm SGRWA}} \left[ \min_{\beta,\lambda} E_\nu^{\rm SGRWA}(\beta,\lambda) \right].
\label{eq:GRWA and VSGRWA_prescription}
\end{gather}

Here $\mathcal R_{\rm GRWA}$ and $\mathcal R_{\rm SGRWA}$ denote the retained eigenvalue branches that can compete for the lowest energy.

% ============================================================
% Observables
% ============================================================

\section{Observables from the Approximate Eigenstates}
\label{sec:4}

The analytical hierarchy developed above provides not only the low-energy spectrum but also
approximate eigenstates, giving direct access to observables that probe the internal structure of the light–matter states of the interacting two-qubit Rabi model. 

Beyond the energy spectrum, we probe this structure through the mean photon number and reduced two-qubit observables. Specifically, we consider the concurrence and equal-axis spin correlations, which together provide complementary tests of the bosonic and qubit sectors of the approximate eigenstates.

% ============================================================
% Photon number
% ============================================================

\subsection{Mean photon number}
\label{subsec:mean_photon_number}

The mean photon number,
\begin{gather}
    \bar n_\nu = \langle\Psi_\nu| \hat a^\dagger\hat a |\Psi_\nu\rangle ,
\label{eq:photon_definition}
\end{gather}
provides a direct probe of the oscillator content of the approximate eigenstates and has also been used to assess displaced variational descriptions of the two-qubit Rabi model~\cite{Mao2019}. Since the analytical states are constructed in transformed frames, the number operator must be evaluated in the corresponding representation. For the displaced adiabatic and GRWA schemes,
\begin{gather}
    \hat U_\beta \hat a^\dagger\hat a \hat U_\beta^\dagger = \hat a^\dagger\hat a - \beta\hat J_x(\hat a^\dagger+\hat a) +
     \beta^2\hat J_x^2 .
\label{eq:number_displaced_operator}
\end{gather}
The three terms describe, respectively, the oscillator occupation in the transformed basis, inter-manifold coherence, and the contribution induced by the spin-dependent displacement. When the lower dressed GRWA branch provides the approximate ground state, its oscillator component is the vacuum and
\begin{gather}
    \bar n_-^{\mathrm{GRWA}} = \frac{\beta^2}{2} \left( 1-\frac{q}{R} \right).
\label{eq:number_GRWA_GS}
\end{gather}
The direct qubit--qubit interactions enter this result through the spin composition of the state. In particular, $q=K/2+\eta_z(A_0-1)/4$ contains $\eta_x$ through $K(\beta)=\omega\beta^2-2g\beta+\eta_x$, whereas $\eta_z$ also modifies the displacement-dependent overlap $A_0$. Thus, the transverse interaction changes the collective-spin contribution to the photon population, while the longitudinal interaction additionally affects the oscillator overlaps and the displacement parameter. For the SGRWA we obtain
\begin{gather}
    \hat W_{\beta,\lambda} \hat a^\dagger\hat a \hat W_{\beta,\lambda}^\dagger = \hat a^\dagger\hat a\cosh(4\lambda) + \sinh^2(2\lambda) \nonumber\\ + \frac{1}{2}\sinh(4\lambda) \left( \hat a^{\dagger2}+\hat a^2 \right) 
    - \beta e^{2\lambda} \hat J_x(\hat a^\dagger+\hat a) + \beta^2\hat J_x^2 .
\label{eq:number_squeezed_operator}
\end{gather}
When the corresponding lower dressed branch provides the approximate ground state, this reduces to
\begin{gather}
    \bar n_-^{\mathrm{SGRWA}} = \sinh^2(2\lambda) + \frac{\beta^2}{2} \left( 1-\frac{q_\lambda}{R_\lambda} \right).
\label{eq:number_SGRWA_GS}
\end{gather}
Squeezing therefore adds the contribution $\sinh^2(2\lambda)$ and modifies the spin-dependent term through the squeezed overlaps. The longitudinal interaction affects both overlaps and the optimized values of $\beta$ and $\lambda$, whereas the transverse interaction still enters mainly through the collective-spin structure.  Explicit branch-dependent expressions are collected in App.~\ref{app:photon_number}.

% ============================================================
% Reduced two-qubit observables
% ============================================================

\subsection{Reduced two-qubit observables}
\label{subsec:reduced_observables}

The energy spectrum and mean photon number provide complementary information about the approximate eigenstates, but they do not determine the quantum state of the two-qubit subsystem. To characterize the material degrees of freedom directly, we therefore consider the reduced state of the two qubits in the ground state. To characterize the two-qubit subsystem, we trace out the bosonic degree of freedom from the selected ground-state eigenvector,
\begin{gather}
    \rho_{12} = \operatorname{Tr}_{\rm ph} \left( |\Psi_{\rm GS}\rangle \langle\Psi_{\rm GS}| \right).
\label{eq:rho12}
\end{gather}
Within the symmetric $j=1$ sector, the ground state can be written as
\begin{gather}
    |\Psi_{\rm GS}\rangle = \sum_n \left( C_{1,n}|1,n\rangle + C_{0,n}|0,n\rangle + C_{-1,n}|-1,n\rangle \right),
\label{eq:GS_spin_expansion}
\end{gather}
where the first label denotes the collective-spin projection and the second the oscillator occupation. We define
\begin{gather}
    P_{+} = \sum_n |C_{1,n}|^2, \; P_{0} = \sum_n |C_{0,n}|^2, \; P_{-} = \sum_n |C_{-1,n}|^2, \nonumber\\ Q =
     \sum_n C_{1,n}C^{*}_{-1,n}.
\label{eq:PQ_definitions}
\end{gather}
Since $|1,0\rangle=(|\uparrow\downarrow\rangle+ |\downarrow\uparrow\rangle)/\sqrt{2}$, the reduced two-qubit state takes the $X$ form
\begin{gather}
    \rho_{12} =
    \begin{pmatrix}
    P_{+} & 0 & 0 & Q\\
    0 & P_{0}/2 & P_{0}/2 & 0\\
    0 & P_{0}/2 & P_{0}/2 & 0\\
    Q^{*} & 0 & 0 & P_{-}
    \end{pmatrix}.
\label{eq:rho12_X}
\end{gather}
Thus, the reduced two-qubit state is completely specified by the three spin-sector populations $P_{\pm,0}$ and the coherence $Q$ between the $m=\pm1$ components. These quantities are obtained from the corresponding GRWA, SGRWA, or numerical ground-state eigenvector and provide a common starting point for the entanglement and spin-correlation observables considered below.

% ============================================================
% Concurrence
% ============================================================

\subsubsection{Concurrence}
\label{subsec:concurrence}

We quantify two-qubit entanglement using Wootters’ concurrence~\cite{Wootters1998}. For the reduced state in Eq.~\eqref{eq:rho12_X}, it takes the compact form
\begin{gather}
C = \max \left\{ 0,\, 2|Q|-P_0,\,
 P_{0}-2\sqrt{P_+P_-} \right\}.
\label{eq:concurrence_PQ}
\end{gather}
Here $C=0$ corresponds to a separable two-qubit state, while $C=1$ represents maximal two-qubit entanglement. After the bosonic mode has been traced out, concurrence measures entanglement between the material degrees of freedom rather than entanglement between the qubits and the field. A change in $C$ therefore signals a change in the reduced two-qubit state, although it does not identify how correlations are distributed in the full light–matter state by itself. In this model, the populations $P_{\pm,0}$ and the coherence $Q$ inherit their dependence on the direct couplings $\eta_x$ and $\eta_z$ from the corresponding ground-state eigenvector. The concurrence therefore probes how correlations generated directly between the qubits combine with those mediated by the common oscillator. This extends the usual two-qubit Rabi setting, where the qubits are correlated only through their coupling to the bosonic mode
~\cite{zhang2015,Lee2013,Chilingaryan2013}.

For each analytical prescription, we evaluate Eq.~\eqref{eq:concurrence_PQ} using the eigenvector of the selected lowest-energy branch. Explicit branch-dependent expressions are collected in App.~\ref{app:lower_reduced_state}.

% ============================================================
% Spin correlations
% ============================================================

\subsubsection{Spin correlations}
\label{subsec:spin_correlations}

Finally, to resolve how the correlations are distributed between different spin directions, we consider the equal-axis correlators
\begin{gather}
    \Gamma_{\alpha\alpha} = \left\langle \hat\sigma_\alpha^{(1)} \hat\sigma_\alpha^{(2)} \right\rangle, \; \Gamma_{\alpha\alpha}^{(c)} = \Gamma_{\alpha\alpha} - \left\langle\hat\sigma_\alpha^{(1)}\right\rangle \left\langle\hat\sigma_\alpha^{(2)}\right\rangle .
\label{eq:spin_correlations_definition}
\end{gather}
For the reduced state in Eq.~\eqref{eq:rho12_X}, the correlations relevant here reduce to
\begin{gather}
    \Gamma_{xx} = P_0+2\operatorname{Re}Q, \quad \Gamma_{zz} = 1-2P_0, \nonumber\\ \Gamma_{zz}^{(c)} = 1-2P_0-(P_+-P_-)^2, \quad \Gamma_{xx}^{(c)}=\Gamma_{xx}.
\label{eq:correlators_PQ}
\end{gather}
Thus, $\Gamma_{xx}$ connects to the coherence between the $m=\pm1$ spin components, while $\Gamma_{zz}$ is determined by the population of the $m=0$ sector. The connected longitudinal correlator removes the part arising from individual-qubit polarizations, leaving only their correlated contribution.

For the lower-dressed GRWA branch, these expressions take the analytical form
\begin{gather}
    \Gamma_{xx}^{(-)} = -\frac{q}{R}, \quad \Gamma_{zz}^{(-)} = \frac{1}{2} \left[ 1+A_0+\frac{q}{R}(1-A_0) \right], \nonumber\\ \Gamma_{zz}^{(c,-)} = \Gamma_{zz}^{(-)} - \left(\frac{h}{R}\right)^2G_0^2 . \label{eq:correlators_lower_branch}
\end{gather}
These expressions make the role of the direct interactions particularly clear. Since
\begin{gather}
    q = \frac{1}{2} \left( \omega\beta^2-2g\beta+\eta_x \right) + \frac{\eta_z}{4}(A_0-1),
\label{eq:q_correlations}
\end{gather}
the transverse coupling $\eta_x$ modifies $\Gamma_{xx}$ directly through the spin mixing encoded in $q/R$. The longitudinal interaction affects the same mixing but also enters $\Gamma_{zz}$ through the displacement-dependent overlap $A_0$. The two interaction channels can therefore redistribute correlations differently between the transverse and longitudinal spin directions. For the squeezed description, the same structure is retained under
\begin{gather}
    G_0\rightarrow G_0^\lambda, \quad A_0\rightarrow A_0^\lambda, \quad \frac{q}{R}\rightarrow\frac{q_\lambda}{R_\lambda}, \quad \frac{h}{R}\rightarrow\frac{h_\lambda}{R_\lambda}.
\label{eq:squeezed_correlator_replacements}
\end{gather}
For other retained ground-state branches, the correlators are evaluated directly from Eq.~\eqref{eq:correlators_PQ}. In the noninteracting limit $\eta_x=\eta_z=0$, these expressions reduce to the known two-qubit Rabi and GRWA results~\cite{zhang2015,Lee2013,Chilingaryan2013}. Explicit branch-dependent expressions are collected in App.~\ref{app:coupled_reduced_state}.

% ============================================================
% Numerical benchmarks
% ============================================================

\section{Numerical Benchmarks and Interaction Effects}
\label{sec:5}

We now compare the analytical approximations with converged numerical diagonalization in the symmetric $j=1$ sector. We first consider the low-energy spectrum and the ground-state energy, then the photon population, and finally the reduced two-qubit observables. This sequence allows us to compare not only the eigenvalues, but also increasingly detailed properties of the corresponding eigenstates. Unless otherwise stated, we work at resonance, $\omega_0/\omega=1$, and use the noninteracting case, $\eta_x=\eta_z=0$, as a reference. We then analyze separately, specific examples: a longitudinal interaction, $\eta_z/\omega=0.5$ with $\eta_x=0$, and a transverse interaction, $\eta_x/\omega=0.5$ with $\eta_z=0$. All energies and coupling strengths
are expressed in units of $\omega$.

% ============================================================
% Spectrum
% ============================================================

\subsection{Spectrum and ground-state accuracy}
\label{subsec:spectrum_accuracy}

\begin{figure*}[t]
\centering
\includegraphics[width=\textwidth]{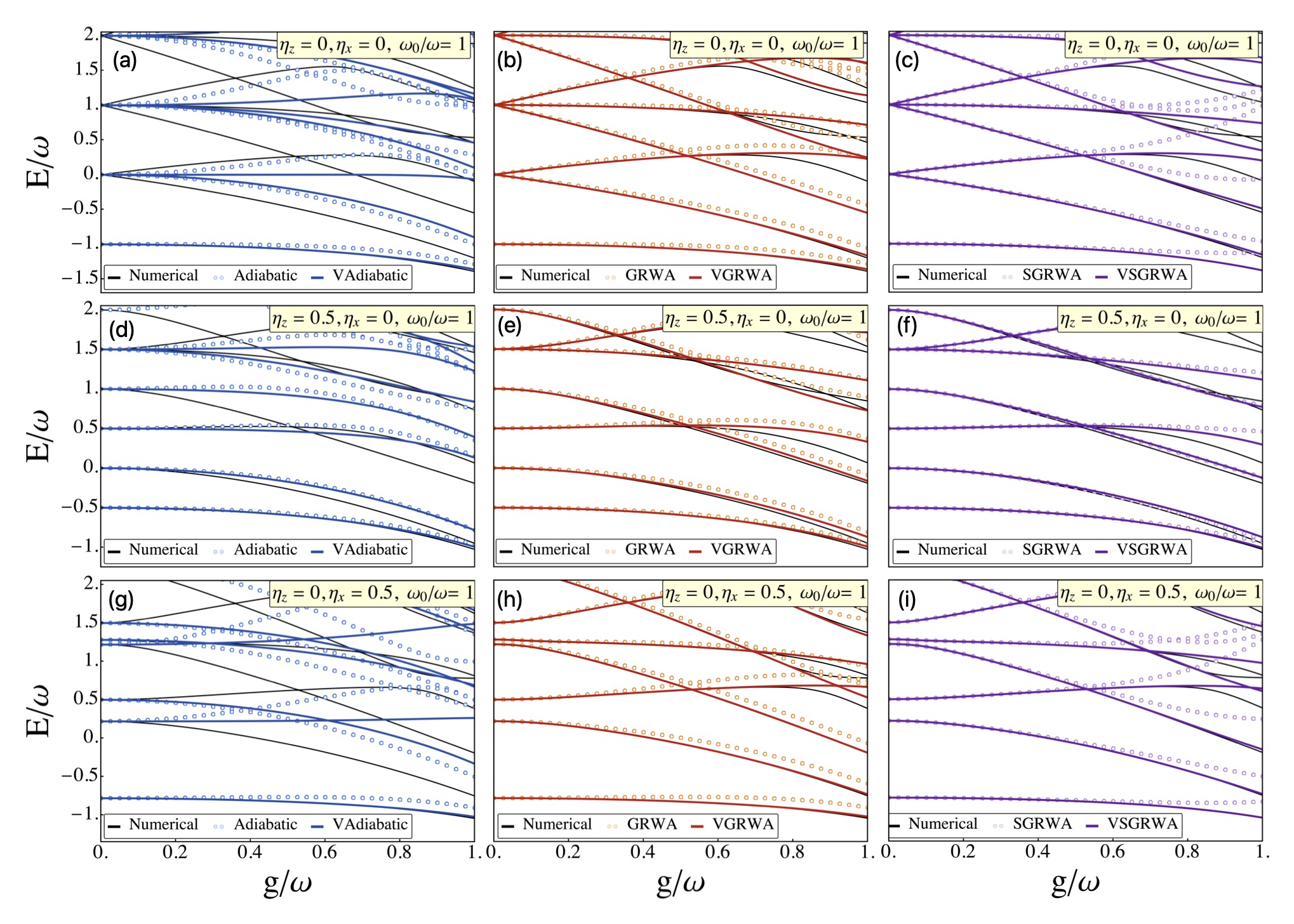}
\caption{
Low-energy spectrum of the interacting two-qubit Rabi model as a function of $g/\omega$ at resonance, $\omega_0/\omega=1$. The upper row corresponds to the noninteracting reference, $\eta_x=\eta_z=0$; the middle row to a longitudinal interaction, $\eta_z/\omega=0.5$ and $\eta_x=0$; and the lower row to a transverse interaction, $\eta_x/\omega=0.5$ and $\eta_z=0$. The left, middle, and right columns show the adiabatic, GRWA, and SGRWA families, respectively. Black solid lines denote numerically exact energies, open colored symbols the fixed-parameter approximations, and solid colored lines their variational counterparts.
}
\label{fig:1}
\end{figure*}

We begin with the low-energy spectrum. Fig.~\ref{fig:1} compares the adiabatic, GRWA, and SGRWA families with numerical diagonalization. The upper row [Figs.~\ref{fig:1} (a)-(c)] gives the noninteracting reference, while the middle [Figs.~\ref{fig:1} (d)-(f)] and lower [Figs.~\ref{fig:1} (g)-(h)] rows show the changes produced by the longitudinal and transverse interactions, respectively.

The resonant regime provides a demanding benchmark because the spin and bosonic energy scales are comparable, so that the low-energy spectrum becomes increasingly sensitive to changes in the bosonic structure and to mixing between oscillator manifolds as $g/\omega$ grows. The interaction-free case allows us to distinguish these conventional light--matter effects from the additional spectral reorganization caused by $\eta_z$ and $\eta_x$. The comparison below is therefore focused on how the two direct interactions modify the spectrum and, consequently, the performance of each level of the analytical hierarchy.

Fig.~\ref{fig:1} first establishes the noninteracting reference against which we can identify the effects of qubit--qubit interactions. In the upper row, the usual behavior of the two-qubit Rabi model is recovered. At small $g/\omega$, all three approximation families describe the low-energy branches reasonably well, whereas differences become progressively more visible in the excited spectrum as the coupling increases. Variational optimization mainly improves the lowest branches, while inter-manifold mixing and squeezing become more important for reproducing the excited-state structure. The middle row shows that the longitudinal interaction substantially reorganizes the spectrum relative to the noninteracting case. In particular, the coupling modifies the positions and separations of the low-energy branches throughout the coupling interval. This behavior is consistent with the structure derived above: $\eta_z$ not only changes the collective-spin energies but also enters the displacement-dependent oscillator matrix elements. Consequently, the longitudinal interaction modifies both the positions of the corresponding energy levels and the amplitudes that drive their hybridization. This distinction is already apparent in the adiabatic comparison, Fig.~\ref{fig:1} (d). Although the approximation still neglects transitions between different displaced oscillator manifolds, the variational result follows the lowest numerical branches closely over a considerable range of $g/\omega$. The improvement therefore originates mainly from adapting the displaced basis to the interaction-modified ground state. Deviations remain for excited branches, as expected from the fixed-manifold character of the approximation.

For a finite longitudinal coupling, $\eta_z$, the relative separation between the low-energy branches changes: some branches move closer while others separate. As $g/\omega$ increases, the points where these branches approach and hybridize shift accordingly. The middle row therefore shows that $\eta_z$ does more than displace the spectrum as a whole; it changes the energy differences that determine the low-energy avoided crossings. The effect of the transverse interaction, $\eta_x$, is different. In this case, the main change comes from the collective-spin composition of the states and from which spin configurations become energetically favorable as the light--matter coupling increases. This is evident in the lower row of Fig.~\ref{fig:1}, where the ordering and evolution of the excited branches differ from both the noninteracting and longitudinal-$\eta_z$ cases. The comparison therefore shows that $\eta_z$ and $\eta_x$ modify the spectrum through different mechanisms, even when both interactions produce visible changes in the positions of the energy levels.

The approximations reproduce this behavior with different levels of accuracy. The variational adiabatic approximation already follows the lowest branches rather well over a broad range of $g/\omega$. This shows that optimizing the displaced bosonic basis captures an important part of the interaction dependence. Its limitations become clearer for the excited states, where the neglected couplings between neighboring manifolds begin to matter. The GRWA improves this part of the spectrum by restoring the dominant couplings between neighboring manifolds. The difference is especially visible when numerical branches approach and hybridize. Squeezing introduces an additional correction, and the VSGRWA gives the most uniform description of the lowest part of the spectrum among the approximations considered. The remaining deviations appear mainly for higher excited states and at larger coupling, where processes outside the retained manifold structure become increasingly important.

The spectrum already shows a clear difference between the ground-state description and the excited-state description. The low-energy structure improves along the analytical hierarchy, but, as expected, the excited branches remain the more demanding part of the comparison. The ground-state energy alone does not show this distinction clearly, so we also consider the normalized absolute error
\begin{gather}
    \frac{\Delta E_{\rm GS}^{X}}{\omega} = \frac{ \left| E_{\rm GS}^{X}-E_{\rm GS}^{\rm num} \right| }{\omega}, \;\; X=\mathrm{VAd,VGRWA,VSGRWA}.
\label{eq:ground_state_error}
\end{gather}

\begin{figure*}[t]
\centering
\includegraphics[width=\textwidth]{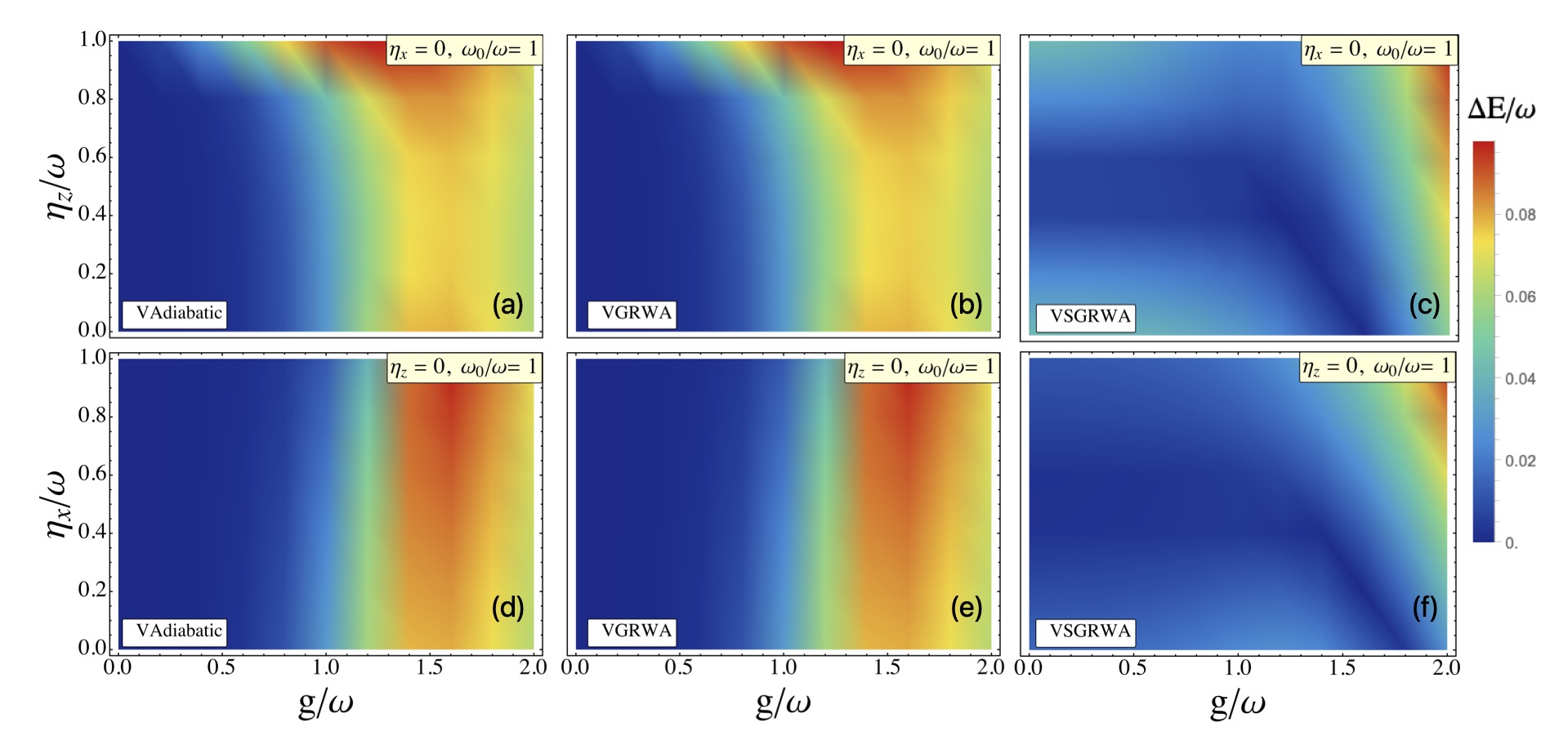}
\caption{
Normalized absolute ground-state energy error $\Delta E_{\rm GS}/\omega$ relative to numerical diagonalization at resonance, $\omega_0/\omega=1$. The upper row scans the longitudinal interaction $\eta_z/\omega$ with $\eta_x=0$, whereas the lower row scans the transverse interaction $\eta_x/\omega$ with $\eta_z=0$. From left to right, the columns correspond to VAdiabatic, VGRWA, and VSGRWA. Lower values of $\Delta E_{\rm GS}/\omega$ indicate better agreement with the numerical ground-state energy.
}
\label{fig:2}
\end{figure*}

Fig.~\ref{fig:2} makes this difference more evident. The VAdiabatic and VGRWA error maps are very similar over broad regions of parameter space, even though Fig.~\ref{fig:1} shows a clear improvement of the excited spectrum when the GRWA couplings are included. Thus, the neighboring-manifold hybridization retained by the GRWA is important for describing the organization of the excited spectrum, but its effect on the ground-state energy can be considerably smaller.

Squeezing has a more visible effect on the ground-state error. The VSGRWA extends the region in which the analytical energy remains close to the numerical result for both interaction channels, although the error increases again at sufficiently large $g/\omega$. The shape of these regions also changes as $\eta_z$ or $\eta_x$ varies. The accuracy of the approximation therefore depends not only on light--matter coupling, but on its competition with the energy scales introduced by qubit--qubit interactions.

% ============================================================
% Photon population
% ============================================================

\subsection{Photon population}
\label{subsec:photon_population}

We now examine the bosonic content of the approximate eigenstates through the mean photon number introduced in Sec.~\ref{sec:4}. This provides information that is not contained in the energies alone, since $\bar n$ depends directly on the amplitudes and coherences of the eigenstates. It therefore gives a more sensitive test of how well the different approximations reproduce their light--matter composition.

\begin{figure*}[t]
    \centering
    \includegraphics[width=\textwidth]{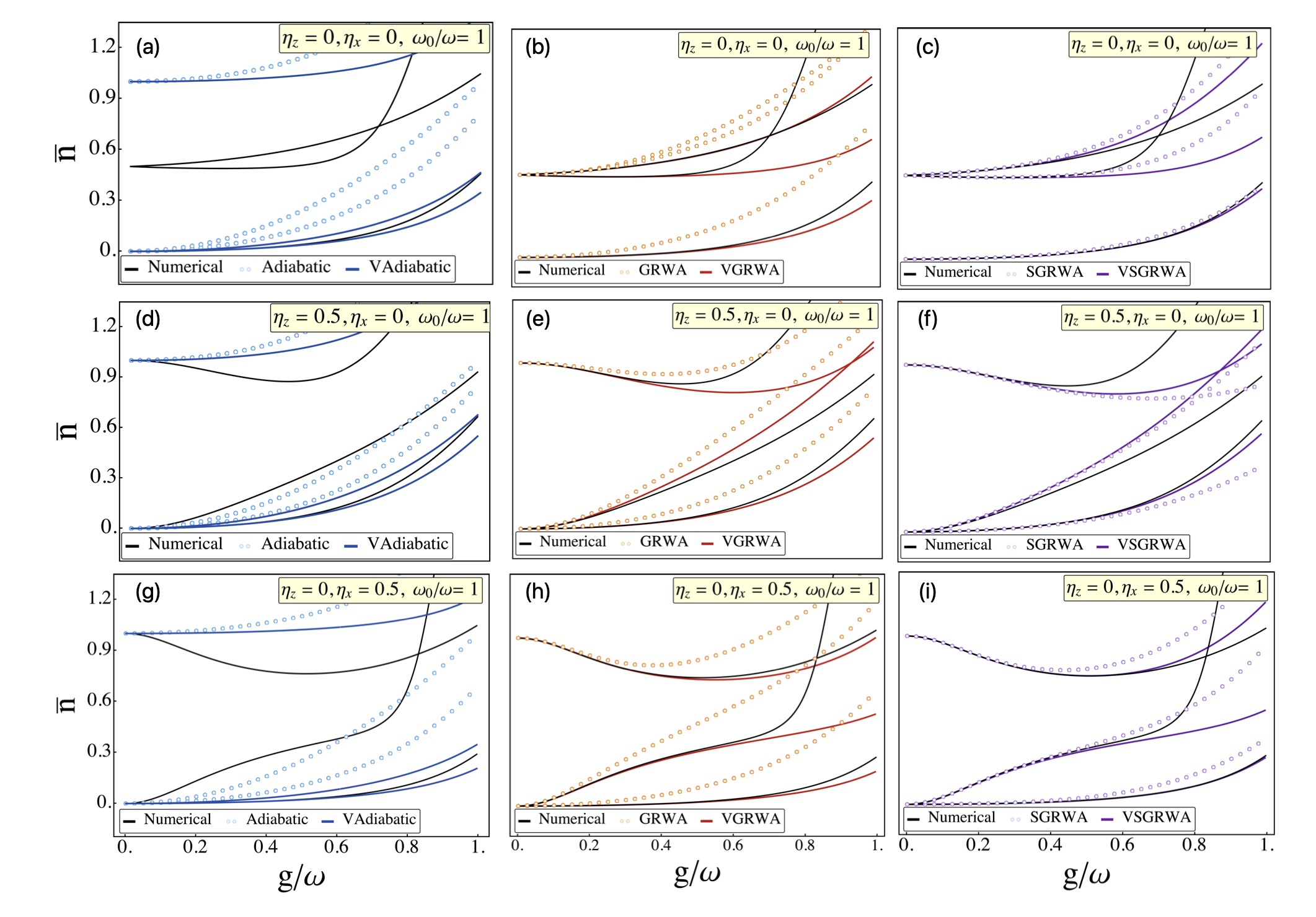}
    \caption{
    Mean photon number $\bar n$ of the ground and low-lying excited states as a function of $g/\omega$ at resonance, $\omega_0/\omega=1$. The upper row corresponds to the noninteracting reference, $\eta_x=\eta_z=0$; the middle row to a longitudinal interaction, $\eta_z/\omega=0.5$ and $\eta_x=0$; and the lower row to a transverse interaction, $\eta_x/\omega=0.5$ and $\eta_z=0$. The left, middle, and right columns show the adiabatic, GRWA, and SGRWA families, respectively. Black solid lines denote numerical results, open colored symbols the fixed-parameter approximations, and solid colored lines their variational counterparts.
    }
    \label{fig:3}
\end{figure*}

The upper row,  Figs.~\ref{fig:3} (a)-(c) provides the noninteracting reference. At resonance, the two lowest excited numerical branches begin close to $\bar n\simeq 1/2$. This behavior arises from hybridization between nearly degenerate states in neighboring oscillator manifolds. The adiabatic approximation excludes this mixing because it treats each fixed-$n$ manifold independently, so its branches start at photon occupations close to integers~\cite{zhang2015}. As $g/\omega$ increases, the lowest branch changes smoothly, while the excited states show a stronger redistribution of their photon population and eventually develop much more rapid variations.

A finite longitudinal interaction, $\eta_z$, changes this picture considerably. In the middle row, Fig.~\ref{fig:3} (d)-(f), one excited branch first loses oscillator population, reaches a pronounced minimum, and then increases rapidly at larger coupling. The other branches do not follow the same evolution and acquire photons at different rates. Thus, the changes produced by $\eta_z$ are not restricted to the energy-level positions discussed above; the oscillator content of the corresponding eigenstates is also reorganized. This is consistent with the approximation’s analytical structure, since $\eta_z$ enters both the displacement-dependent overlaps and the parameters that define the displaced and squeezed bosonic bases. The lower row, Figs.~\ref{fig:3} (g)-(i), shows a different behavior when the transverse interaction $\eta_x$ is present. Unlike $\eta_z$, this interaction does not generate additional displacement-dependent matrix elements, but it changes the collective-spin composition of the eigenstates. The different spin components therefore couple to the oscillator with different weights as $g/\omega$ is varied. In this way, $\eta_x$ can substantially redistribute the photon population even though its effect originates mainly in the qubit sector.

The differences between the approximations become clearer for the excited states. The variational adiabatic approximation follows the smooth ground-state photon population reasonably well, but its independent manifolds cannot reproduce all the changes produced by inter-manifold mixing. The GRWA recovers an important part of this behavior by coupling neighboring manifolds, while the SGRWA also allows the oscillator fluctuations to adjust through squeezing. Both optimized approximations describe the lowest branches well over a broad coupling range. Their largest deviations occur where the numerical photon populations change rapidly, which signals a stronger redistribution of the components of the corresponding eigenstates.

The photon population therefore makes visible differences that are much less evident from the energies. An approximation can reproduce the ground-state energy accurately without reproducing the state’s oscillator composition with the same precision. This distinction is particularly important for the excited branches, which provide a more demanding test of the manifold structure retained by the GRWA and SGRWA. We now turn to the ground-state reduced two-qubit state, where the same approximate eigenvectors can be tested through concurrence and spin correlations.

% ============================================================
% Correlation and entanglement
% ============================================================

\subsection{Concurrence and two-qubit correlations}
\label{subsec:two_qubit_correlations}

We finally examine the ground-state two-qubit correlations through the concurrence and spin observables introduced in Sec.~\ref{subsec:reduced_observables}. These quantities probe the reduced two-qubit density matrix and therefore provide a more demanding test of the approximate eigenstates than the energy or a single bosonic expectation value.

\begin{figure*}[t]
    \centering
    \includegraphics[width=0.78\textwidth]{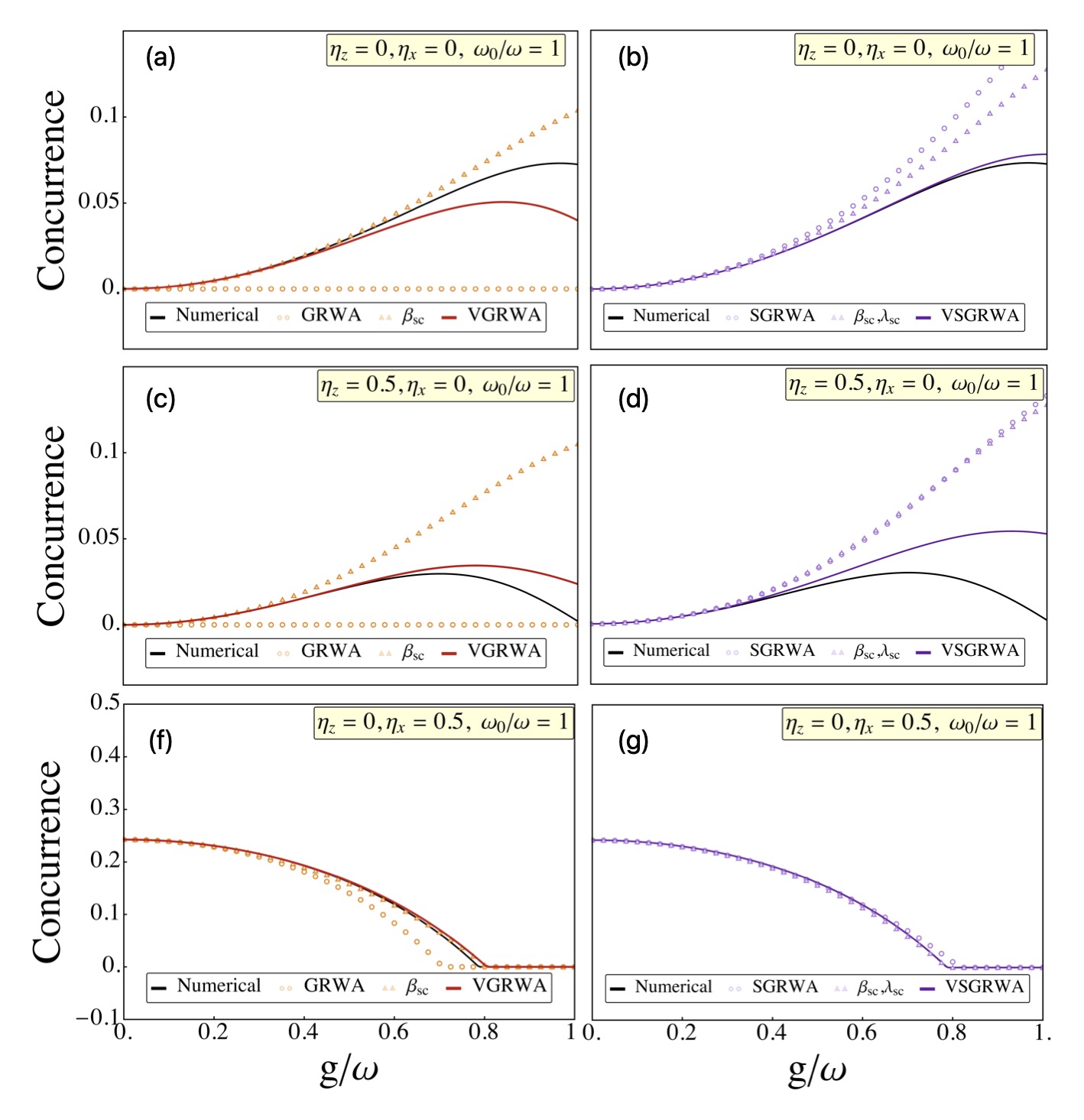}
    \caption{
    Ground-state two-qubit concurrence as a function of $g/\omega$ at resonance, $\omega_0/\omega=1$. The upper row corresponds to the noninteracting reference, $\eta_x=\eta_z=0$; the middle row to a longitudinal interaction, $\eta_z/\omega=0.5$ and $\eta_x=0$; and the lower row to a transverse interaction, $\eta_x/\omega=0.5$ and $\eta_z=0$. The left column compares the displaced GRWA prescriptions, while the right column shows their squeezed counterparts. Black solid lines denote numerical results; open symbols denote the fixed and self-consistent prescriptions, and solid colored lines denote the fully variational approximations.
    }
    \label{fig:4}
\end{figure*}

The upper row, Figs.~\ref{fig:4} (a)-(b), shows the entanglement generated in the absence of a direct qubit--qubit interaction. Starting from zero at $g=0$, the concurrence becomes small but finite as the light--matter coupling increases. This matches previous results for the two-qubit Rabi model, where counter-rotating processes generate effective correlations between the qubits via virtual photons, and the concurrence grows quadratically at weak coupling~\cite{Lee2013}. In Figs.~\ref{fig:4} (c)-(d), a finite longitudinal interaction, $\eta_z$, keeps this oscillator-mediated origin at weak coupling, but changes its subsequent evolution: the concurrence first increases, reaches a maximum, and then decreases again at larger $g/\omega$. The situation changes when the transverse interaction $\eta_x$ is present, as observed in Figs.~\ref{fig:4} (f)-(g). Because this term directly mixes the two-qubit configurations, the ground state is already appreciably entangled at $g=0$. Increasing the light--matter coupling then reduces this pairwise entanglement until the concurrence vanishes. In this case, the oscillator is not simply generating correlations between initially separable qubits; it competes with the correlations already favored by the direct transverse interaction.

The different approximations follow these two behaviors with variable accuracy. For $\eta_x\neq0$, both VGRWA and VSGRWA closely reproduce the decrease and eventual disappearance of the concurrence very closely. The longitudinal case is more demanding. The variational approximations recover the initial growth and the turnover position, but deviations remain after the maximum, especially for VSGRWA. This shows that the reduced two-qubit state is more sensitive to $\eta_z$ than the ground-state energy or photon population alone.

\begin{figure*}[t]
    \centering
    \includegraphics[width=0.82\textwidth]{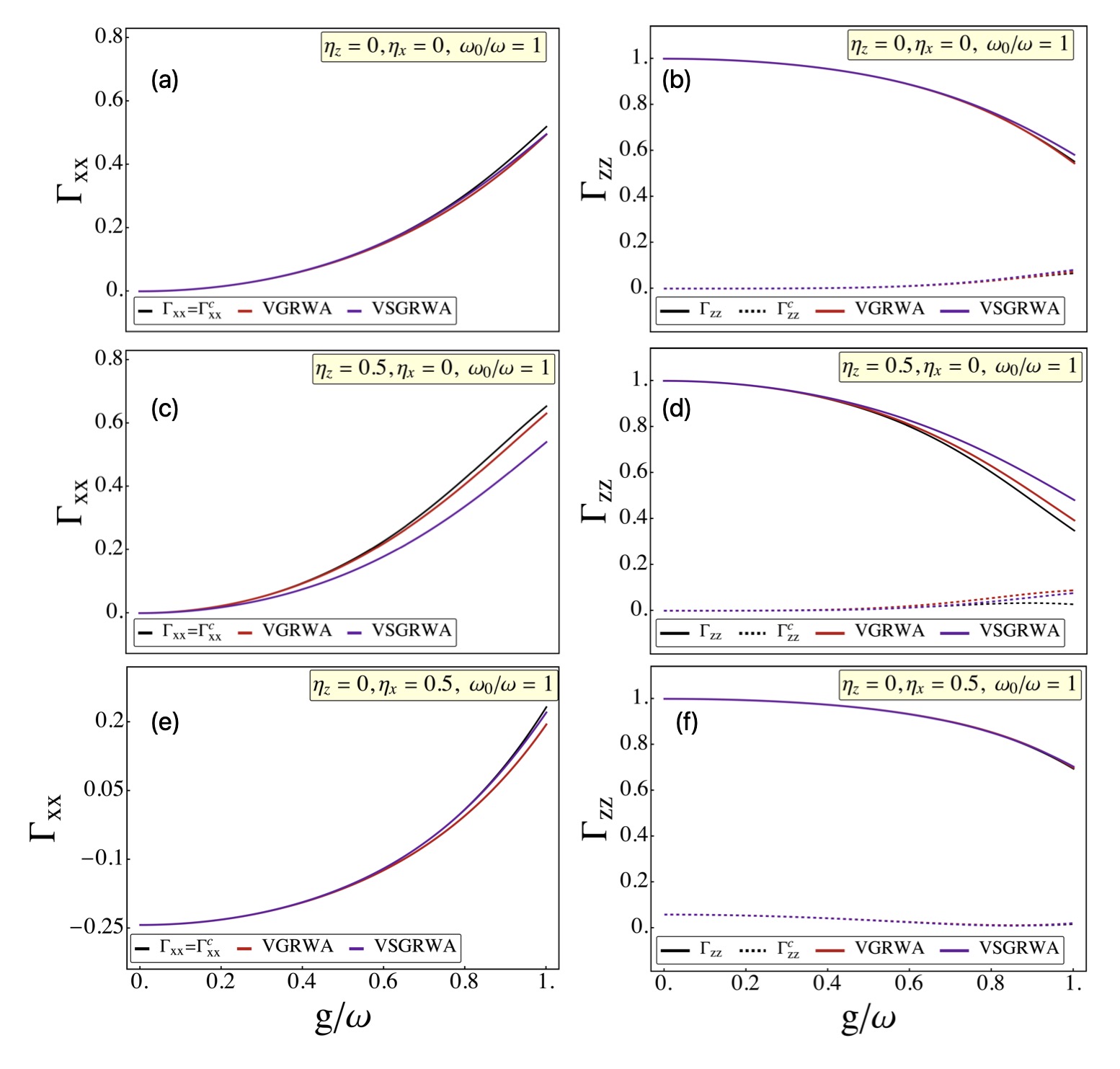}
    \caption{
    Ground-state equal-axis spin correlations as a function of $g/\omega$ at resonance, $\omega_0/\omega=1$. The upper, middle, and lower rows correspond respectively to $\eta_x=\eta_z=0$, $\eta_z/\omega=0.5$ with $\eta_x=0$, and $\eta_x/\omega=0.5$ with $\eta_z=0$. The left column shows $\Gamma_{xx}$, for which $\Gamma_{xx}^{(c)}=\Gamma_{xx}$, while the right column shows $\Gamma_{zz}$ and its connected component $\Gamma_{zz}^{(c)}$. Black curves denote numerical results, and red and purple curves correspond to VGRWA and VSGRWA, respectively; dotted curves denote the connected longitudinal correlator.
    }
    \label{fig:5}
\end{figure*}

The spin correlations in Fig.~\ref{fig:5} show how this reorganization is distributed between the transverse and longitudinal directions. In the noninteracting case, increasing $g/\omega$ produces positive transverse correlations $\Gamma_{xx}$ while $\Gamma_{zz}$ decreases. Adding a finite $\eta_z$, as seen in Fig.~\ref{fig:5} (c)-(d), strengthens this redistribution: $\Gamma_{xx}$ increases more rapidly, and the longitudinal correlation rapidly decreases. This is also the case where the difference between the two optimized approximations becomes more visible, with VGRWA staying closer to the numerical correlations at larger coupling than VSGRWA.

From Fig.~\ref{fig:5} (e)-(f), the transverse interaction gives a different and particularly clear signature. For finite $\eta_x$, $\Gamma_{xx}$ is initially negative and changes sign as $g/\omega$ increases. The correlations favored directly by $\eta_x$ are therefore progressively replaced by correlations of opposite sign as the coupling to the common oscillator becomes stronger. At the same time, $\Gamma_{zz}$ remains large over the interval shown and is reproduced accurately by both VGRWA and VSGRWA.

% ============================================================
% Summary and conclusions
% ============================================================
\section{Circuit-QED implementation}
\label{circuit::qed}
Whether the model of Eq.~\eqref{eq:interacting_two_qubit_qrm} can be realized at all in the
ultrastrong or the deep-strong regimes hinges on the choice of corresponding artificial atom.
A light-matter coupling comparable to the qubit and oscillator frequencies demands a large interaction and threatens the two-level description on which
the model rests, two requirements that most superconducting qubits cannot meet
simultaneously. The artificial atoms of our proposed circuit implementation are therefore
three-junction persistent-current (flux) qubits \cite{Orlando1999,Mooij1999}. Since their properties determine which terms of Eq.~\eqref{eq:interacting_two_qubit_qrm} can
be realized and at which strength, we summarize here how such a loop behaves as a two-level system and why it is the appropriate choice for the considered regimes.

\begin{figure*}[t]
    \centering
    \includegraphics[width=0.85\textwidth]{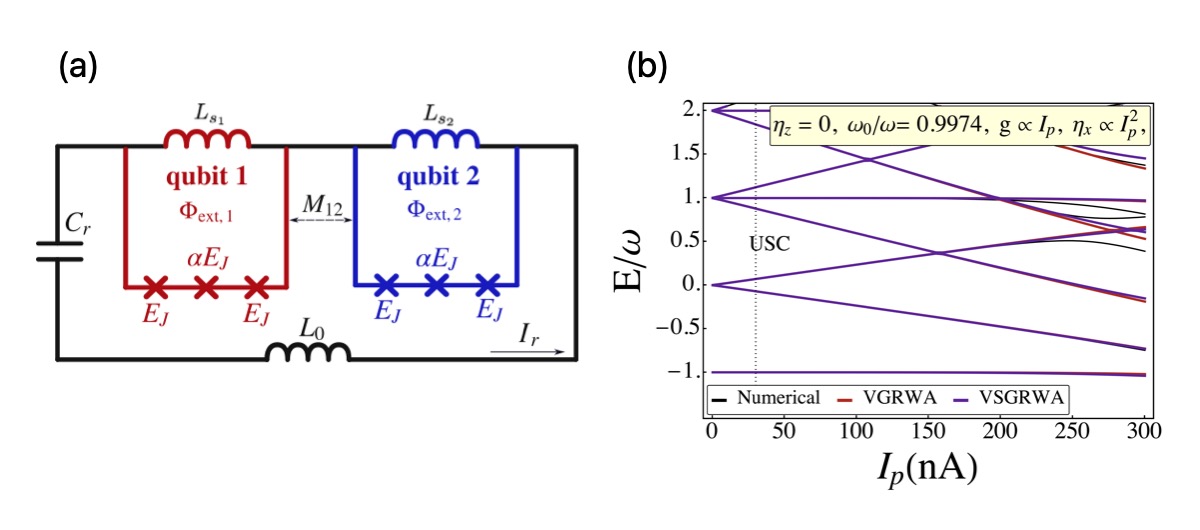}
    \caption{Circuit-QED implementation of the interacting two-qubit Rabi model. (a) Two three-junction flux qubits coupled to a common LC resonator through separate segments $L_{s1}$ and $L_{s2}$. The mutual inductance $M_{12}$ provides the direct qubit--qubit interaction. (b) Low-energy spectrum of the interacting two-qubit Rabi model as a function of the persistent current $I_p$, with $\omega_0/\omega=0.9974$ and $\eta_z=0$. The couplings $g$ and $\eta_x$ depend on $I_p$ according to Eq.~\eqref{eq:current_dependent_couplings}. Black lines show the numerical results, while red and purple lines show VGRWA and VSGRWA, respectively. The vertical black dotted line at $I_p\simeq30.06\,\mathrm{nA}$ indicates $g/(\hbar\omega)=0.1$ using the collective coupling parameter.}
    \label{fig:circuit_current_spectrum}
\end{figure*}

\subsection{The loop as a qubit.}

Figure~\ref{fig:circuit_current_spectrum} (a) shows the proposed circuit, which consists of two identical three-junction flux qubits coupled to a common LC resonator. Each qubit has two Josephson junctions with energy $E_J$ and a smaller junction with energy $\alpha E_J$. The external fluxes $\Phi_{\mathrm{ext},1}$ and $\Phi_{\mathrm{ext},2}$ control the bias of each qubit. For each superconducting loop threaded by a flux $\Phi$, fluxoid quantization imposes a constraint on the junction phases, leaving only two independent phases. At half a
flux quantum and for $\alpha>1/2$ it has two degenerate minima at
$\varphi_-=\pm\varphi^{*}$, $\cos\varphi^{*}=1/2\alpha$, in which the loop
carries opposite circulating currents $\pm I_p$,
$I_p=I_c\sqrt{1-1/4\alpha^{2}}$, screening the flux toward $0$ or $\Phi_0$.
The small junction sets the barrier between them; the junction capacitances
give the phases a mass, so the two current states tunnel into each other with
amplitude $\Delta/2$.
Projected onto the two
lowest levels, with Pauli operators $\hat{\tau}_i$ in the current basis,
\begin{equation}
\hat H_q=\frac{\varepsilon}{2}\hat{\tau}_z+\frac{\Delta}{2}\hat{\tau}_x,\qquad
\hat I_q=I_p\,\hat{\tau}_z,\qquad \varepsilon=2I_p\Big(\Phi-\frac{\Phi_0}{2}\Big),
\label{eq:fq_TLS}
\end{equation}
where $\hat I_q$ is the circulating-current operator. The loop flux sets
$\varepsilon$; replacing the small junction by a dc-SQUID makes $\Delta$
tunable with a second flux line \cite{Paauw2009}.  At the symmetry point
($\varepsilon=0$) the eigenstates are $(|{+I_p}\rangle\pm|{-I_p}\rangle)/\sqrt2$
with splitting $\hbar\omega_0=\Delta$, and $\hat I_q$ is purely off-diagonal
between them.

\subsection{Interactions}

Because the only operator through which the loop interacts with its
surroundings is $\hat I_q$, every coupling is set by a mutual inductance. Each qubit loop is closed through its
own segment $L_s$ of the oscillator inductor, so the oscillator current
$\hat I_r=I_{\rm zpf}(\hat a+\hat a^\dagger)$, $I_{\rm zpf}=\sqrt{\hbar\omega/2L_r}$,
threads the loop with the flux $L_s\hat I_r$.  Expanding the magnetic energy
$\tfrac12L_s(\hat I_r+\hat I_q)^2$ gives the coupling $L_s\hat I_r\hat I_q$,
i.e.\ a mutual inductance equal to the shared self-inductance, which is why
galvanic sharing reaches couplings that geometric mutual inductances cannot
\cite{Bourassa2009,Niemczyk2010,Yoshihara2017}; the qubit self-term
$\tfrac12L_s\hat I_q^2=\tfrac12L_sI_p^2$ is a constant, so no diamagnetic
($A^2$) term appears.  A small mutual inductance $M_{12}$ between the two
loops adds $M_{12}\hat I_{q1}\hat I_{q2}$.  Rotating to the energy eigenbasis
($\hat{\tau}_x\to\hat{\sigma}_z$, $\hat{\tau}_z\to\hat{\sigma}_x$) and setting $\varepsilon_j=0$ yields
Eq.~(\ref{eq:interacting_two_qubit_qrm}) with \begin{equation}
\hbar\omega_0=\Delta,\qquad
\frac{g}{2}=L_sI_pI_{\rm zpf},\qquad
\frac{\eta_x}{2}=M_{12}I_p^{2},\qquad
\eta_z\simeq0 .
\label{eq:fq_map}
\end{equation}
The two qubits must sit on \emph{different} segments of the inductor: a common
segment would add $L_sI_p^2\hat{\sigma}_x^1\hat{\sigma}_x^2$, i.e.\ $\eta_x\sim(I_p/I_{\rm zpf})g\sim30g$,
whereas $M_{12}$ of a few pH gives $\eta_x\sim\hbar\omega/2$ independently of
$g$.  The $\eta_z$ term is absent because $\hat I_q$ is transverse; the only
longitudinal observable, the state-dependent current of the $\alpha$-SQUID, is
$\sim10^2$ smaller, and biasing away from $\varepsilon=0$ would add a
longitudinal oscillator coupling of relative size $\sqrt{\eta_z/\eta_x}$.

Three properties make this design the right one at $g\sim\hbar\omega$.
The qubit states are the lowest doublet of a double well, with the next level
at the plasma frequency $\sim\sqrt{8E_JE_C}/\hbar$, tens of GHz away, so the
two-level truncation survives couplings that would excite the third and
fourth levels of a transmon.  The coupling operator $I_p\sim100$--$500$~nA is
an order of magnitude larger than a transmon's effective dipole, which together
with galvanic coupling gives $g/\hbar\omega$ of order unity
\cite{Yoshihara2017}.  And since the double well is defined by
Josephson rather than geometric inductance, the loop is only micrometers in
size: the fluxoid description holds, the coupling is a clean mutual
inductance, and flux-noise sensitivity vanishes to first order at the symmetry
point.  The price is that $\Delta$ depends exponentially on $\alpha$ and
$E_J/E_C$ and that coherence degrades away from $\varepsilon=0$; tunable-gap
\cite{Paauw2009} and capacitively shunted \cite{Yan2016} variants mitigate
this at the cost of a smaller $I_p$.  For deep-strong coupling, where $I_p$
and anharmonicity matter most, the original design remains the one used in most
experiments to date \cite{Yoshihara2017,Yoshihara2018}. Some representative parameters are given in Table \ref{table::params}.

\begin{table}[t]
\caption{Representative parameters.}
\label{tab:appparams}
\begin{ruledtabular}
\begin{tabular}{lll}
Quantity & Value & Experimental parameters of Eq. \eqref{eq:interacting_two_qubit_qrm} \\
\hline
$L_{r} $, $C_{r} $ & 6 nH, 1.05 pF & $\omega/2\pi=2$ GHz, $I_{\mathrm{zpf}} =10.5$ nA \\
$\alpha$, $E_{J} /E_{C} $ & $0.6$--$0.7$, $30$--$50$ & $\Delta/h=2$ GHz, $I_{p} \simeq300$ nA \\
$L_{s} $ & $0\to0.21$ nH & $g/\hbar\omega=0\to1$ \\
$M_{12}$ & $0$ or $3.7$ pH & $\eta_x=0$ or $\hbar\omega/2$ \\
\end{tabular}
\end{ruledtabular}
\label{table::params}
\end{table}

To connect the circuit parameters in Table~\ref{table::params} with the results of Sec.~\ref{sec:5}, we study how the energy spectrum changes with the  current $I_p$. We keep $L_s=0.21\,\mathrm{nH}$ and $M_{12}=3.7\,\mathrm{pH}$ constant, while $I_p$ changes from $0$ to $300\,\mathrm{nA}$. In this case, both the light--matter coupling $g$ and the transverse qubit--qubit interaction $\eta_x$ change with the current. Using the relations obtained above and the resonator parameters $L_r=6\,\mathrm{nH}$ and $C_r=1.05\,\mathrm{pF}$, we obtain

\begin{gather}
\frac{g(I_p)}{\hbar\omega}=0.997877\frac{I_p}{300\,\mathrm{nA}}, \notag\\
\frac{\eta_x(I_p)}{\hbar\omega}=0.501266\left(\frac{I_p}{300\,\mathrm{nA}}\right)^2.
\label{eq:current_dependent_couplings}
\end{gather}

Figure~\ref{fig:circuit_current_spectrum} (b) shows the energy spectrum obtained from numerical diagonalization, VGRWA, and VSGRWA. As the current increases, $g$ grows linearly, while $\eta_x$ grows quadratically. As a result both interactions change the energy levels at the same time. The two variational methods reproduce the lowest energy levels well. In this calculation we keep the qubit frequency $\omega_0$ constant.

\section{Summary and Conclusions}
\label{sec:6}

We have studied the interacting two-qubit Rabi model in the symmetric spin-one sector using the adiabatic, GRWA, and SGRWA approximations, including fixed, self-consistent, and variational choices of the bosonic parameters. The direct qubit--qubit interactions can be incorporated within the same analytical framework without changing the basic hierarchy of the approximations. The adiabatic approximation retains independent displaced manifolds, the GRWA restores the dominant couplings between neighboring manifolds, and the SGRWA additionally allows the oscillator fluctuations to adjust through squeezing. The different prescriptions for the displacement ($\beta$) and squeezing ($\lambda$) parameters then determine how each approximation adapts the bosonic basis.

The effects of $\eta_x$ and $\eta_z$ are clearly different. The transverse interaction $\eta_x$ mainly changes the collective-spin composition of the states and competes directly with the correlations generated through the common oscillator. For the parameters considered here, the ground state is already entangled at $g=0$, while increasing $g/\omega$ progressively suppresses this pairwise entanglement. The accompanying sign change in $\Gamma_{xx}$ clearly signals this competition. The longitudinal interaction $\eta_z$ affects both the qubit and oscillator parts of the problem: it modifies the displacement-dependent overlaps and generates the additional two-boson matrix element $W_n$ in the interacting GRWA. This difference is reflected in the spectrum, photon population, and reduced two-qubit state, where finite $\eta_z$ causes the ground-state concurrence to increase and subsequently decrease.

The numerical comparison also shows that the accuracy of an approximation's accuracy depends on the quantity considered. Variational optimization accurately describes the lowest energies over a broad parameter range, whereas the neighboring-manifold couplings retained in the GRWA become particularly important for the excited-state spectrum. Squeezing further enlarges the region of quantitative ground-state accuracy and improves several bosonic properties. The photon population is nevertheless more sensitive to the eigenstate structure than the energy alone, and concurrence and spin correlations provide an even stronger test of the reduced two-qubit state. This is particularly evident for finite $\eta_z$, where VGRWA and VSGRWA can remain accurate for energetic and bosonic quantities while showing larger differences in the two-qubit observables. Thus, increasing the flexibility of the bosonic basis does not necessarily improve every property of the ground state by the same amount.

These results show that displaced and squeezed analytical approximations remain useful for the interacting two-qubit Rabi model beyond the RWA, while preserving a direct connection between the retained matrix elements and the physical processes they describe. The largest discrepancies occur at stronger coupling and in higher excited states, where additional oscillator manifolds and multiphoton processes become increasingly relevant. Extending the retained manifold structure is therefore a natural continuation of the present analysis, together with asymmetric qubit configurations and multimode extensions. 

Furthermore, we have shown that tunable qubit-qubit interactions are experimentally available in the two-qubit Rabi model via circuit QED implementations, paving the way toward exploring the hierarchy of approximations in practical platforms relevant in the quantum information context.

\begin{acknowledgments} 
R. H. R. acknowledges support from the Graduate Program scholarship from CONAHCYT/Secihti. M. A. B. M. acknowledges financial support from CONAHCYT/Secihti No. CBF2023-2024-1765, the PEAPDI 2026 project from the CBI UAM-I, and the Marcos Moshinsky Fellowship. C. A. G--G. acknowledges  SECIHTI--Mexico for financyal support under grant No. CBF2023-2024-2888, as well as DGAPA-PAPIIT-UNAM under grant No. IA104625.
\end{acknowledgments} 

%%%%%%%%%%%%%%%%%%%%%%%%%%%%%%%%%%%%%%%%
\section*{Data Availability}
%%%%%%%%%%%%%%%%%%%%%%%%%%%%%%%%%%%%%%%%
The data that support the findings of this article are openly available~\cite{}.

%%%
%%%
%%%
\appendix

% ============================================================
% Matrix elements 
% ============================================================

\section{Matrix elements and analytical details}
\label{app:analytical_details}

This Appendix collects the matrix elements and intermediate steps required to reconstruct the analytical approximations introduced in Sec.~\ref{sec:3}. We focus on the ingredients that are not explicit in the main text: the displaced-oscillator overlaps, the origin of the retained GRWA couplings, their squeezed extension, and the self-consistent bosonic parameters.

% ============================================================
% Displaced-oscillator matrix elements
% ============================================================

\subsection{Displaced-oscillator matrix elements}
\label{app:displaced_matrix_elements}

The displaced approximations are generated by matrix elements of

\begin{gather}
    \hat D(\alpha) = \exp\!\left[ \alpha(\hat a^\dagger-\hat a) \right].
\label{eq:app_displacement_operator}
\end{gather}

For real $\alpha$ and $r\geq0$,

\begin{gather}
    \langle n|\hat D(\alpha)|n+r\rangle = (-\alpha)^r \sqrt{\frac{n!}{(n+r)!}}\, e^{-\alpha^2/2} L_n^{(r)}(\alpha^2).
\label{eq:app_displacement_matrix}
\end{gather}
The diagonal matrix element is therefore even in $\alpha$, whereas off-diagonal elements with an odd change of oscillator occupation change sign under $\alpha\rightarrow-\alpha$. This immediately gives

\begin{gather}
    G_n = \langle n| \cosh[\beta(\hat a^\dagger-\hat a)] |n\rangle = e^{-\beta^2/2}L_n(\beta^2), \\ A_n = \langle n| \cosh[2\beta(\hat a^\dagger-\hat a)] |n\rangle = e^{-2\beta^2}L_n(4\beta^2),
\label{eq:app_diagonal_overlaps}
\end{gather}
together with the one-boson factors
\begin{gather}
    -\langle n| \sinh[\beta(\hat a^\dagger-\hat a)] |n+1\rangle = R_n \nonumber \\ = \frac{\beta}{\sqrt{n+1}} e^{-\beta^2/2} L_n^{(1)}(\beta^2), \\ -\langle n| \sinh[2\beta(\hat a^\dagger-\hat a)] |n+1\rangle = B_n \nonumber \\ = \frac{2\beta}{\sqrt{n+1}} e^{-2\beta^2} L_n^{(1)}(4\beta^2).
\label{eq:app_one_boson_overlaps}
\end{gather}
The even matrix element connecting oscillator states that differ by two is

\begin{gather}
    A_n^{(2)} = \langle n-1| \cosh[2\beta(\hat a^\dagger-\hat a)] |n+1\rangle \nonumber \\ = \frac{4\beta^2}{\sqrt{n(n+1)}} e^{-2\beta^2} L_{n-1}^{(2)}(4\beta^2).
\label{eq:app_two_boson_overlap}
\end{gather}

% ============================================================
% GRWA
% ============================================================

\subsection{Retained GRWA matrix elements}
\label{app:grwa_matrix_elements}

The GRWA uses the dressed collective-spin basis defined by the lowest oscillator manifold in Sec.~\ref{subsec:grwa}. When this same basis is used for an arbitrary manifold $n$, the diagonal spin Hamiltonian is not exactly diagonal. The remaining coupling between the $|-\rangle$ and $|+\rangle$ dressed states is

\begin{gather}
    \Lambda_n = \langle -|\hat H_s^{(n)}|+\rangle = q_n\cos\theta-h_n\sin\theta,
\label{eq:app_Lambda_n}
\end{gather}

with

\begin{gather}
    \Lambda_0=0.
\label{eq:app_Lambda_zero}
\end{gather}
Thus, the reference manifold is diagonal by definitions, while $\Lambda_n$ measures the residual same-manifold mixing for $n>0$. This term is neglected in the GRWA, whereas the dominant transitions between neighboring oscillator manifolds are retained.

The couplings between neighboring oscillator manifolds are obtained by projecting the displaced Hamiltonian of Eq.~\eqref{eq:Hbeta} between $|n\rangle$ and $|n+1\rangle$.

\begin{gather}
    \hat T_n \equiv \langle n+1|\hat H(\beta)|n\rangle .
\label{eq:app_Tn_definition}
\end{gather}
For a one-boson transition, the diagonal oscillator terms do not contribute, while the residual linear coupling and the terms containing the odd displacement overlaps are retained. Using Eq.~\eqref{eq:app_one_boson_overlaps}, this gives

\begin{gather}
    \hat T_n = (g-\omega\beta)\sqrt{n+1}\,\hat J_x - \omega_0R_n(-i\hat J_y) \nonumber \\
    - \frac{\eta_z}{2}B_n \left[-i\{\hat J_z,\hat J_y\}\right].
\label{eq:app_Tn}
\end{gather}
The two one-boson couplings entering the GRWA block are consequently

\begin{gather}
    Z_n = \langle -,n+1|\hat T_n|0,n\rangle, \; Y_n = \langle 0,n+1|\hat T_n|+,n\rangle,
\label{eq:app_ZY_def}
\end{gather}

whose explicit forms are Eqs.~\eqref{eq:Zn} and~\eqref{eq:Yn} of the main text.

A distinct contribution arises from the longitudinal interaction because the even part of the transformed $\hat J_z^2$ term can change the oscillator occupation by two. Defining

\begin{gather}
    \hat K_2 = \frac{\hat J_z^2-\hat J_y^2}{2},
\label{eq:app_K2}
\end{gather}

the spin-one matrices give

\begin{gather}
    \hat J_z^2=
    \begin{pmatrix}
    1&0&0\\
    0&0&0\\
    0&0&1
    \end{pmatrix},
    \qquad
    \hat J_y^2=
    \frac{1}{2}
    \begin{pmatrix}
    1&0&-1\\
    0&2&0\\
    -1&0&1
    \end{pmatrix}.
\label{eq:app_JzJy_squared}
\end{gather}

Hence, in the $\{|1\rangle,|-1\rangle\}$ subspace,

\begin{gather}
    \hat K_2^{(\pm1)}
    =
    \frac{1}{4}
    \begin{pmatrix}
    1&1\\
    1&1
    \end{pmatrix}.
\label{eq:app_K2_matrix}
\end{gather}

Using $|-\rangle=-s|1\rangle+c|-1\rangle$ and $|+\rangle=c|1\rangle+s|-1\rangle$, with $c=\cos(\theta/2)$ and $s=\sin(\theta/2)$, one obtains

\begin{gather}
    \langle-|\hat K_2|+\rangle = \frac{1}{4}(c-s)(c+s) = \frac{1}{4}\cos\theta.
\label{eq:app_K2_dressed}
\end{gather}
Combining this spin matrix element with the two-boson overlap $A_n^{(2)}$ gives

\begin{gather}
    W_n = \eta_z A_n^{(2)} \langle-|\hat K_2|+\rangle = \frac{\eta_z}{4}A_n^{(2)}\cos\theta.
\label{eq:app_Wn}
\end{gather}

Thus, $W_n$ is generated specifically by the longitudinal interaction and vanishes for $\eta_z=0$.

After these truncations, the effective $3\times3$ Hamiltonian is the block given in Eq.~\eqref{eq:HGRWA}. For completeness, its characteristic polynomial can be written as

\begin{gather}
    \det\!\left[ E\mathbb I-\hat H_n^{\rm GRWA} \right] = E^3-T_nE^2+S_nE-P_n = 0,
\label{eq:app_GRWA_cubic}
\end{gather}

where

\begin{gather}
    T_n = V_{-,n}+V_{0,n}+V_{+,n},
    \label{eq:app_T_cubic}
    \\
    S_n = V_{-,n}V_{0,n} + V_{-,n}V_{+,n} + V_{0,n}V_{+,n} \nonumber \\
    - Z_n^2 - Y_{n-1}^2 - W_n^2,
    \label{eq:app_S_cubic}
    \\
    P_n = V_{-,n}V_{0,n}V_{+,n} + 2Z_nY_{n-1}W_n - V_{-,n}Y_{n-1}^2 \nonumber \\
    - V_{0,n}W_n^2 - V_{+,n}Z_n^2.
\label{eq:app_P_cubic}
\end{gather}

The three GRWA energies are the roots of Eq.~\eqref{eq:app_GRWA_cubic}. Following the Cardano form used in related GRWA treatments, such as Ref.~\cite{zhang2015}, we define

\begin{gather}
    \Delta_n=T_n^2-3S_n, \; \Theta_n= \frac{1}{3} \arccos\!\left[ \frac{ 2T_n^3-9T_nS_n+27P_n }{ 2\Delta_n^{3/2} } \right].
\label{eq:app_cardano_parameters}
\end{gather}
The three analytical eigenvalues are then
\begin{gather}
E_{n,k}^{\rm GRWA} = \frac{T_n}{3} + \frac{2}{3}\sqrt{\Delta_n} \cos\!\left( \Theta_n-\frac{2\pi k}{3} \right), \; k=0,1,2.
\label{eq:app_GRWA_energies}
\end{gather}

This expression provides the closed analytical spectrum of each retained GRWA block and is the form used in our calculations.

% ============================================================
% SGWRA
% ============================================================

\subsection{Squeezed extension}
\label{app:squeezed_extension}

The SGRWA follows the same truncation structure after applying the squeezing transformation introduced in Sec.~\ref{subsec:sgrwa}. Its action on the oscillator operators is

\begin{gather}
    \hat V_\lambda\hat a\hat V_\lambda^\dagger = \hat a\cosh(2\lambda) + \hat a^\dagger\sinh(2\lambda), \\ \hat V_\lambda\hat a^\dagger\hat V_\lambda^\dagger = \hat a^\dagger\cosh(2\lambda) + \hat a\sinh(2\lambda).
\label{eq:app_squeeze_a}
\end{gather}

Hence,

\begin{gather}
    \hat V_\lambda (\hat a^\dagger+\hat a) \hat V_\lambda^\dagger = e^{2\lambda} (\hat a^\dagger+\hat a),  \\ \hat V_\lambda (\hat a^\dagger-\hat a) \hat V_\lambda^\dagger = e^{-2\lambda} (\hat a^\dagger-\hat a).
\label{eq:app_squeeze_quadratures}
\end{gather}
The displacement appearing in all oscillator overlaps is therefore renormalized according to

\begin{gather}
    \bar\beta = \beta e^{-2\lambda},
\label{eq:app_beta_bar}
\end{gather}

which directly gives

\begin{gather}
    \left\{
    G_n^\lambda,
    A_n^\lambda,
    R_n^\lambda,
    B_n^\lambda,
    A_n^{(2),\lambda}
    \right\}
    \nonumber \\=
    \left.
    \left\{
    G_n,
    A_n,
    R_n,
    B_n,
    A_n^{(2)}
    \right\}
    \right|_{\beta\rightarrow\bar\beta}.
\label{eq:app_squeezed_overlaps}
\end{gather}

The oscillator Hamiltonian transforms as

\begin{gather}
    \omega\hat V_\lambda \hat a^\dagger\hat a \hat V_\lambda^\dagger = \Omega_\lambda\hat a^\dagger\hat a + E_\lambda + \frac{\omega}{2}\sinh(4\lambda) \left( \hat a^{\dagger2}+\hat a^2 \right),
\label{eq:app_squeezed_oscillator}
\end{gather}
with
\begin{gather}
    E_\lambda = \omega\sinh^2(2\lambda), \qquad \Omega_\lambda = \omega\cosh(4\lambda).
\label{eq:app_EOmega_lambda}
\end{gather}
The explicit pair-creation and pair-annihilation term in Eq.~\eqref{eq:app_squeezed_oscillator} is neglected within the SGRWA truncation. In contrast, the two-boson contribution proportional to $\eta_z$ is retained because it originates from the transformed qubit--qubit interaction and belongs to the extended GRWA connectivity.

Because the dressed-spin basis is defined by the $n=0$ manifold, it does not exactly diagonalize the spin block for $n>0$. The remaining same-manifold coupling is therefore

\begin{gather}
    \Lambda_n^\lambda \equiv {}_\lambda\langle-| \hat H_s^{(n)} |+\rangle_\lambda = q_n^\lambda\cos\theta_\lambda - h_n^\lambda\sin\theta_\lambda, \;\Lambda_0^\lambda=0,
\label{eq:app_Lambda_lambda}
\end{gather}

and is neglected within the SGRWA truncation.

The coefficient $\Gamma_n^\lambda$ instead originates from the residual linear term $(g-\omega\beta)\hat J_x(\hat a^\dagger+\hat a)$. Since squeezing gives $\hat a^\dagger+\hat a\rightarrow e^{2\lambda}(\hat a^\dagger+\hat a)$ and $\langle n+1|\hat a^\dagger+\hat a|n\rangle=\sqrt{n+1}$, its one-boson amplitude is

\begin{gather}
    \Gamma_n^\lambda = (g-\omega\beta)e^{2\lambda}\sqrt{n+1}.
\label{eq:app_Gamma_lambda}
\end{gather}
Together with the squeezed overlaps of Eq.~\eqref{eq:app_squeezed_overlaps}, these quantities generate $Z_n^\lambda$, $Y_n^\lambda$, and $W_n^\lambda$ of Eqs.~\eqref{eq:Zlambda}--\eqref{eq:Wlambda}.

% ============================================================
% Parameteres
% ============================================================

\subsection{Self-consistent bosonic parameters}
\label{app:self_consistent_parameters}

The fixed and fully variational prescriptions are defined directly in Sec.~\ref{subsec:parameter_prescriptions}. Here we give the short derivation of the intermediate self-consistent parameters. Following the displaced GRWA approximation of Irish~\cite{irish2007} and its squeezed extension by Zhang~\cite{zhangyu2016}, the bosonic basis is optimized at the reference state level before diagonalizing the retained GRWA or SGRWA blocks.

For the displaced basis, the corresponding reference-state energy is

\begin{gather}
    E_{\rm DB}(\beta) = -\omega_0e^{-\beta^2/2} + \frac{1}{2} \left[ (\omega\beta-2g)\beta+\eta_x \right] \nonumber \\ + \frac{\eta_z}{4}e^{-2\beta^2} + \frac{3\eta_z}{4}.
\label{eq:app_EDB}
\end{gather}

Stationarity gives

\begin{gather}
    \frac{dE_{\rm DB}}{d\beta} = \beta \left[ \omega + \omega_0e^{-\beta^2/2} - \eta_ze^{-2\beta^2} \right] - g = 0,
\label{eq:app_EDB_stationary}
\end{gather}

which immediately leads to the self-consistent displacement Eq.~\eqref{eq:beta_sc}.

When squeezing is included, the analogous reference-state energy is

\begin{gather}
    E_{\rm SDB}(\beta,\lambda) = \omega\sinh^2(2\lambda) - \omega_0e^{-\bar\beta^2/2} \nonumber \\ + \frac{1}{2} \left[ (\omega\beta-2g)\beta+\eta_x \right] 
    + \frac{\eta_z}{4}e^{-2\bar\beta^2} + \frac{3\eta_z}{4},
\label{eq:app_ESDB}
\end{gather}

where $\bar\beta=\beta e^{-2\lambda}$. The two stationary conditions can be written compactly as

\begin{gather}
    0 = \frac{\partial E_{\rm SDB}}{\partial\beta} = \omega\beta-g + \beta e^{-4\lambda} \left[ \omega_0e^{-\bar\beta^2/2}
    - \eta_ze^{-2\bar\beta^2} \right], \label{eq:app_ESDB_beta}
    \\
    0 = \frac{\partial E_{\rm SDB}}{\partial\lambda} = 2\omega\sinh(4\lambda) - 2\bar\beta^2 \left[ \omega_0e^{-\bar\beta^2/2} - \eta_ze^{-2\bar\beta^2} \right].
\label{eq:app_ESDB_lambda}
\end{gather}

Rearranging Eqs.~\eqref{eq:app_ESDB_beta} and \eqref{eq:app_ESDB_lambda} gives the coupled self-consistent equations \eqref{eq:sgrwa_sc_beta} and~\eqref{eq:sgrwa_sc_lambda} quoted in the main text. Expanding them for $\beta^2\ll1$ and $|\lambda|\ll1$ yields the noniterative estimates in Eq.~\eqref{eq:approx_squeezed_parameters}.

The self-consistent approximation therefore changes only the bosonic basis used to evaluate the retained blocks. The fully variational schemes go one step further by minimizing the actual GRWA or SGRWA eigenvalue branches, as defined in Sec.~\ref{subsec:parameter_prescriptions}.

% ============================================================
% Numerical diagonalization
% ============================================================

\section{Numerical diagonalization}
\label{app:numerical}

The numerical reference used throughout Sec.~\ref{sec:5} is obtained by direct diagonalization of the Hamiltonian in the symmetric $j=1$ sector. This Appendix specifies the truncated basis, matrix representation and observables.

We omit the constant shift $E_S^0=-(\eta_x+\eta_z)/2$, as in the
analytical treatment, and diagonalize

\begin{gather}
    \hat H = \omega\hat a^\dagger\hat a + \omega_0\hat J_z +
    g\hat J_x(\hat a^\dagger+\hat a) + \eta_x\hat J_x^2 + \eta_z\hat J_z^2 .
\label{eq:app_num_H}
\end{gather}

% ============================================================
% Hilbert spaces
% ============================================================

\subsection{Truncated Hilbert space}
\label{app:num_basis}

The oscillator Fock space is truncated at $n=n_{\max}$, while the complete spin-one sector is retained. The numerical basis is therefore

\begin{gather}
    \mathcal{B}_{\rm num} = \left\{ |1,m\rangle\otimes|n\rangle: m=1,0,-1,\; n=0,\ldots,n_{\max} \right\},
\label{eq:app_num_basis}
\end{gather}

with dimension

\begin{gather}
    D_{\rm num} = 3(n_{\max}+1).
\label{eq:app_num_dimension}
\end{gather}

We use the spin ordering $\{|1,1\rangle,|1,0\rangle,|1,-1\rangle\}$ and the oscillator ordering $\{|0\rangle,|1\rangle,\ldots,|n_{\max}\rangle\}$.

For the oscillator sector,

\begin{gather}
    a = \sum_{n=1}^{n_{\max}} \sqrt{n}\, |n-1\rangle\langle n|,
    \; N=a^\dagger a = \operatorname{diag}(0,1,\ldots,n_{\max}),
\label{eq:app_oscillator_matrices}
\end{gather}

and we define

\begin{gather}
    X=a^\dagger+a.
\label{eq:app_num_X_definition}
\end{gather}

Defining $N_F=n_{\max}+1$, the Hamiltonian in the symmetric $j=1$ sector is represented in the truncated basis as

\begin{gather}
    H_{\rm num} = \omega I_3\otimes N + \omega_0J_z\otimes I_{N_F} + gJ_x\otimes X  \nonumber \\
    +\eta_xJ_x^2\otimes I_{N_F} + \eta_zJ_z^2\otimes I_{N_F},
\label{eq:app_numerical_matrix}
\end{gather}

where $I_3$ and $I_{N_F}$ denote the identity operators in the spin and truncated oscillator spaces, respectively.

In the spin ordering specified above, this representation takes the explicit block form
\begin{widetext}
\begin{gather}
    H_{\rm num} =
    \begin{pmatrix}
    \omega N+
    \left(
    \omega_0+\eta_z+\frac{\eta_x}{2}
    \right)I_{N_F}
    &
    \frac{g}{\sqrt{2}}X
    &
    \frac{\eta_x}{2}I_{N_F}
    \\[2mm]
    \frac{g}{\sqrt{2}}X
    &
    \omega N+\eta_x I_{N_F}
    &
    \frac{g}{\sqrt{2}}X
    \\[2mm]
    \frac{\eta_x}{2}I_{N_F}
    &
    \frac{g}{\sqrt{2}}X
    &
    \omega N+
    \left(
    -\omega_0+\eta_z+\frac{\eta_x}{2}
    \right)I_{N_F}
    \end{pmatrix}.
\label{eq:app_numerical_block_matrix}
\end{gather}
\end{widetext}
Each entry is an $N_F\times N_F$ oscillator-space matrix, so that $H_{\rm num}$ has total dimension $3N_F=3(n_{\max}+1)$. Equation~\eqref{eq:app_numerical_block_matrix} is the matrix diagonalized for each parameter set $(g,\omega,\omega_0,\eta_x,\eta_z)$.

% ============================================================
% Eigenstates
% ============================================================

\subsection{Eigenstates and observables}
\label{app:num_observables}

The numerical eigenvalue problem is

\begin{gather}
H_{\rm num}\mathbf C^{(r)} = E_r^{\rm num}\mathbf C^{(r)}, \; E_0^{\rm num}\leq E_1^{\rm num}\leq\cdots ,
\label{eq:app_num_eigenproblem}
\end{gather}

where $r$ labels the ordered eigenstates. The corresponding normalized state is

\begin{gather}
    |\Psi_r^{\rm num}\rangle = \sum_{m=-1}^{1} \sum_{n=0}^{n_{\max}} C_{m,n}^{(r)} |1,m\rangle\otimes|n\rangle, \;\sum_{m,n}|C_{m,n}^{(r)}|^2=1.
\label{eq:app_num_state}
\end{gather}

The mean photon number follows directly from the same eigenvector,

\begin{gather}
    \bar n_r^{\rm num} = \sum_{m=-1}^{1} \sum_{n=0}^{n_{\max}} n\,|C_{m,n}^{(r)}|^2.
\label{eq:app_num_photon}
\end{gather}

For the ground-state concurrence and spin correlations, the coefficients $C_{m,n}^{(0)}$ are used to construct $P_+$, $P_0$, $P_-$, and $Q$ according to Sec.~\ref{subsec:reduced_observables}. These quantities completely determine the reduced two-qubit density matrix and, therefore, the concurrence and spin correlations defined there. Thus, energies, photon populations, concurrence, and spin correlations are all evaluated from the same numerical eigenvectors.

% ============================================================
% Observables
% ============================================================

\section{Observables from the approximate eigenstates}
\label{app:approx_observables}

This Appendix gives the expressions needed to evaluate the observables of Sec.~\ref{sec:4} from the analytical eigenstates. We first provide compact photon-number formulas for the adiabatic, GRWA, and SGRWA states, and then derive the reduced two-qubit state for the lower dressed branch. A general prescription is finally given when the approximate ground state belongs to a coupled GRWA or SGRWA block.

% ============================================================
% Photon number 
% ============================================================
\subsection{Photon number}
\label{app:photon_number}

For the displaced schemes, the number operator in the transformed frame is

\begin{gather}
    \hat n_{\beta} =
    \hat U_{\beta}\hat a^\dagger\hat a\hat U_{\beta}^\dagger = \hat a^\dagger\hat a - \beta\hat J_x(\hat a^\dagger+\hat a) + \beta^2\hat J_x^2 .
\label{eq:app_number_displaced}
\end{gather}

For an adiabatic eigenstate belonging to a single oscillator manifold $n$, the term linear in $\hat a^\dagger+\hat a$ has vanishing expectation value. Defining

\begin{gather}
    q_n^{\rm Ad} = K(\beta) + \frac{\eta_z}{2} \left[A_n(\beta)-1\right], \\ R_n^{\rm Ad} = \sqrt{ \left(q_n^{\rm Ad}\right)^2 + 4\omega_0^2G_n^2(\beta) },
\label{eq:app_ad_photon_parameters}
\end{gather}

the three adiabatic branches give

\begin{gather}
    \bar n_{n,0}^{\rm Ad} = n+\beta^2, \quad \bar n_{n,\pm}^{\rm Ad} = n+ \frac{\beta^2}{2} \left( 1\pm \frac{q_n^{\rm Ad}}{R_n^{\rm Ad}} \right).
\label{eq:app_ad_photon}
\end{gather}

The fixed and variational adiabatic results follow from the corresponding choice of $\beta$.

For the GRWA, let

\begin{gather}
    |\widetilde{\Psi}_{n,j}^{\rm GRWA}\rangle = C_-^{(n,j)}|-,n+1\rangle + C_0^{(n,j)}|0,n\rangle \nonumber \\ + C_+^{(n,j)}|+,n-1\rangle,
\label{eq:app_grwa_state}
\end{gather}

with $n\geq1$. Where

\begin{gather}
    \mathbf C_{n,j} =
    \begin{pmatrix}
    C_-^{(n,j)}\\
    C_0^{(n,j)}\\
    C_+^{(n,j)}
    \end{pmatrix},
    \qquad
    \mathbf C_{n,j}^\dagger\mathbf C_{n,j}=1.
\label{eq:app_grwa_vector}
\end{gather}

Introducing

\begin{gather}
    \chi=\frac{q}{R}, \qquad c=\cos\frac{\theta}{2}, \qquad s=\sin\frac{\theta}{2},
\label{eq:app_grwa_photon_parameters}
\end{gather}

the photon population can be written compactly as

\begin{gather}
    \bar n_{n,j}^{\rm GRWA} = \mathbf C_{n,j}^\dagger \mathcal N_n^{\rm GRWA} \mathbf C_{n,j},
\label{eq:app_grwa_photon_matrix_form}
\end{gather}

with
\begin{widetext}
\begin{gather}
    \mathcal N_n^{\rm GRWA} =
    \begin{pmatrix}
    n+1+\dfrac{\beta^2}{2}(1-\chi)
    &
    -\beta\sqrt{\dfrac{n+1}{2}}(c-s)
    &
    0
    \\[3mm]
    -\beta\sqrt{\dfrac{n+1}{2}}(c-s)
    &
    n+\beta^2
    &
    -\beta\sqrt{\dfrac{n}{2}}(c+s)
    \\[3mm]
    0
    &
    -\beta\sqrt{\dfrac{n}{2}}(c+s)
    &
    n-1+\dfrac{\beta^2}{2}(1+\chi)
    \end{pmatrix}.
\label{eq:app_grwa_photon_matrix}
\end{gather}
\end{widetext}
The diagonal terms contain the oscillator occupation and the spin-dependent displacement, while the off-diagonal elements account for the coherence between neighboring components of the GRWA eigenvector.

For the exceptional $n=0$ block, we write

\begin{gather}
    |\widetilde{\Psi}_{0,j}^{\rm GRWA}\rangle = C_0^{(j)}|0,0\rangle + C_-^{(j)}|-,1\rangle .
\label{eq:app_grwa_zero_state}
\end{gather}

In the ordered basis $\{|0,0\rangle,|-,1\rangle\}$,

\begin{gather}
    \mathcal N_0^{\rm GRWA}
    =
    \begin{pmatrix}
    \beta^2
    &
    -\dfrac{\beta}{\sqrt{2}}(c-s)
    \\[2mm]
    -\dfrac{\beta}{\sqrt{2}}(c-s)
    &
    1+\dfrac{\beta^2}{2}(1-\chi)
    \end{pmatrix}.
\label{eq:app_grwa_zero_number}
\end{gather}

The decoupled lower branch $|-,0\rangle$ is recovered directly as

\begin{gather}
    \bar n_-^{\rm GRWA} = \frac{\beta^2}{2}(1-\chi).
\label{eq:app_grwa_lower_number}
\end{gather}

For the squeezed approximation,

\begin{gather}
    \hat n_{\beta,\lambda} = \hat W_{\beta,\lambda} \hat a^\dagger\hat a \hat W_{\beta,\lambda}^\dagger = \hat a^\dagger\hat a\cosh(4\lambda) + \sinh^2(2\lambda) \nonumber \\ 
    + \frac{1}{2}\sinh(4\lambda) \left( \hat a^{\dagger2}+\hat a^2 \right)
    - \beta e^{2\lambda} \hat J_x(\hat a^\dagger+\hat a) + \beta^2\hat J_x^2 .
\label{eq:app_number_squeezed}
\end{gather}

Within the retained SGRWA manifold, the anomalous term $\hat a^{\dagger2}+\hat a^2$ has zero expectation value: states whose oscillator occupations differ by two carry orthogonal dressed-spin components.

Defining

\begin{gather}
    \kappa_\lambda=\cosh(4\lambda), \quad n_{\rm sq}=\sinh^2(2\lambda), \quad \chi_\lambda=\frac{q_\lambda}{R_\lambda},
\label{eq:app_sgrwa_number_parameters}
\end{gather}

and

$c_\lambda=\cos(\theta_\lambda/2)$, $s_\lambda=\sin(\theta_\lambda/2)$, the photon population for $n\geq1$ takes the same quadratic form,

\begin{gather}
    \bar n_{n,j}^{\rm SGRWA} = (\mathbf C_{n,j}^{\lambda})^\dagger \mathcal N_n^{\rm SGRWA} \mathbf C_{n,j}^{\lambda},
\label{eq:app_sgrwa_photon_matrix_form}
\end{gather}

where
\begin{widetext}
\begin{gather}
    \mathcal N_n^{\rm SGRWA} =
    \begin{pmatrix}
    \kappa_\lambda(n+1)+n_{\rm sq}
    +\dfrac{\beta^2}{2}(1-\chi_\lambda)
    &
    -\beta e^{2\lambda}
    \sqrt{\dfrac{n+1}{2}}
    (c_\lambda-s_\lambda)
    &
    0
    \\[3mm]
    -\beta e^{2\lambda}
    \sqrt{\dfrac{n+1}{2}}
    (c_\lambda-s_\lambda)
    &
    \kappa_\lambda n+n_{\rm sq}+\beta^2
    &
    -\beta e^{2\lambda}
    \sqrt{\dfrac{n}{2}}
    (c_\lambda+s_\lambda)
    \\[3mm]
    0
    &
    -\beta e^{2\lambda}
    \sqrt{\dfrac{n}{2}}
    (c_\lambda+s_\lambda)
    &
    \kappa_\lambda(n-1)+n_{\rm sq}
    +\dfrac{\beta^2}{2}(1+\chi_\lambda)
    \end{pmatrix}.
\label{eq:app_sgrwa_photon_matrix}
\end{gather}
\end{widetext}
For the exceptional squeezed block,

\begin{gather}
    \mathcal N_0^{\rm SGRWA} =
    \begin{pmatrix}
    n_{\rm sq}+\beta^2
    &
    -\dfrac{\beta e^{2\lambda}}{\sqrt{2}}
    (c_\lambda-s_\lambda)
    \\[2mm]
    -\dfrac{\beta e^{2\lambda}}{\sqrt{2}}
    (c_\lambda-s_\lambda)
    &
    \kappa_\lambda+n_{\rm sq}
    +\dfrac{\beta^2}{2}(1-\chi_\lambda)
\end{pmatrix}.
\label{eq:app_sgrwa_zero_number}
\end{gather}

The squeezed lower branch consequently gives

\begin{gather}
    \bar n_-^{\rm SGRWA} =
    \sinh^2(2\lambda) + \frac{\beta^2}{2} (1-\chi_\lambda),
\label{eq:app_sgrwa_lower_number}
\end{gather}

as quoted in the main text.

% ============================================================
% Concurrence
% ============================================================

\subsection{Reduced state of the lower dressed branch}
\label{app:lower_reduced_state}

When the lower dressed GRWA branch is the selected ground state, its
transformed-frame state is

\begin{gather} |\widetilde G_-\rangle
  =|0\rangle_{\rm ph}\otimes|-\rangle, \qquad |G_-\rangle = \hat U_\beta^\dagger|\widetilde G_-\rangle.
\label{eq:app_lower_state}
\end{gather}

Writing the dressed spin state in the $\hat J_x$ basis gives

\begin{gather}
    |G_-\rangle = u|-\beta\rangle|+_x\rangle + v|0\rangle|0_x\rangle + u|\beta\rangle|-_x\rangle,
\label{eq:app_lower_coherent_state}
\end{gather}

where

\begin{gather}
    u=\frac{c-s}{2}, \qquad v=\frac{c+s}{\sqrt{2}}.
\label{eq:app_uv}
\end{gather}

The different collective-spin components are therefore correlated with
different coherent states of the oscillator. Their overlap is controlled
by

\begin{gather}
    G=e^{-\beta^2/2},
    \qquad
    A=e^{-2\beta^2}.
\label{eq:app_GA}
\end{gather}

For compactness, define

\begin{gather}
    \chi=\frac{q}{R}, \qquad \zeta=\frac{h}{R}, \qquad \chi^2+\zeta^2=1.
\label{eq:app_chi_zeta}
\end{gather}

After tracing out the oscillator and returning to the two-qubit basis $\{|\uparrow\uparrow\rangle,|\uparrow\downarrow\rangle, |\downarrow\uparrow\rangle,|\downarrow\downarrow\rangle\}$, the reduced state has the X form

\begin{gather}
    \rho_{12}^{(-)} =
    \begin{pmatrix}
    \rho_{11} & 0 & 0 & \rho_{14}\\
    0 & \rho_{22} & \rho_{22} & 0\\
    0 & \rho_{22} & \rho_{22} & 0\\
    \rho_{14} & 0 & 0 & \rho_{44}
    \end{pmatrix},
\label{eq:app_rho_lower}
\end{gather}

with

\begin{gather}
    \rho_{11} = \frac{ 3+A+\chi(1-A)-4\zeta G }{8},
    \\
    \rho_{44} = \frac{ 3+A+\chi(1-A)+4\zeta G }{8},
    \label{eq:app_rho_diag}
    \\
    \rho_{14} = \frac{ A(1-\chi)-1-3\chi }{8},  \\ \rho_{22} = \frac{ (1-\chi)(1-A) }{8}.
\label{eq:app_rho_offdiag}
\end{gather}

These expressions make explicit the two ingredients entering the reduced two-qubit state: $\chi$ and $\zeta$ determine its dressed-spin composition, while $G$ and $A$ quantify the remaining coherence between oscillator components associated with different spin states.

Using the X-state concurrence of Sec.~\ref{sec:4}, the lower branch gives

\begin{gather}
    \mathcal C_- = \max\left\{ 0,\, \mathcal C_1,\, \mathcal C_2 \right\},
\label{eq:app_concurrence_lower}
\end{gather}

where

\begin{gather}
    \mathcal C_1 = \frac{1}{4} \left[ \left| A(1-\chi)-1-3\chi \right| - (1-\chi)(1-A) \right],
    \label{eq:app_concurrence_C1}
    \\ \mathcal C_2 = \frac{1}{4} \left[ (1-\chi)(1-A) \right. \nonumber \\
    \left. - \sqrt{ \left[ 3+A+\chi(1-A) \right]^2 - 16G^2(1-\chi^2) } \right].
\label{eq:app_concurrence_C2}
\end{gather}

The corresponding equal-axis correlations reduce to

\begin{gather}
    \Gamma_{xx}^{(-)} = -\chi, \qquad \Gamma_{zz}^{(-)} = \frac{1}{2} \left[ 1+A+\chi(1-A) \right],
\label{eq:app_lower_correlations}
\end{gather}

and, since $\langle\sigma_z^{(1)}\rangle = \langle\sigma_z^{(2)}\rangle = -\zeta G$,

\begin{gather}
    \Gamma_{zz}^{(c,-)} = \Gamma_{zz}^{(-)} - \zeta^2G^2, \qquad \Gamma_{xx}^{(c,-)} = \Gamma_{xx}^{(-)}.
\label{eq:app_lower_connected}
\end{gather}

The squeezed lower branch has exactly the same reduced-state structure. One only replaces

\begin{gather}
    \beta\rightarrow\bar\beta=\beta e^{-2\lambda}, \; G\rightarrow G_\lambda=e^{-\bar\beta^2/2}, \; A\rightarrow A_\lambda=e^{-2\bar\beta^2}, \nonumber\\ \chi\rightarrow \chi_\lambda=\frac{q_\lambda}{R_\lambda}, \quad \zeta\rightarrow \zeta_\lambda=\frac{h_\lambda}{R_\lambda}.
\label{eq:app_squeezed_reduced_replacements}
\end{gather}

Equations~\eqref{eq:app_rho_lower}--\eqref{eq:app_lower_connected} therefore apply directly to the squeezed description after these substitutions.

% ============================================================
% Correlations
% ============================================================

\subsection{Reduced state for coupled GRWA and SGRWA branches}
\label{app:coupled_reduced_state}

If the lowest retained energy belongs to a coupled block, the reduced state must be constructed from its complete eigenvector rather than from the lower dressed branch alone. For a GRWA eigenvector Eq.~\eqref{eq:app_grwa_state}, let

\begin{gather} 
    \nu_- = n+1, \qquad \nu_0=n, \qquad \nu_+=n-1,
\label{eq:app_nu_alpha}
\end{gather}

and denote the dressed-spin states collectively by $|\alpha\rangle$, with $\alpha=-,0,+$. Expanding them in the eigenbasis $|\mu_x\rangle$ of $\hat J_x$, $\mu=-1,0,1$, the physical coefficients in the $\hat J_z\otimes$Fock basis are

\begin{gather}
    \mathcal C_{m,\ell}^{(n,j)} = \nonumber \\
    \sum_{\alpha=-,0,+} C_\alpha^{(n,j)} \sum_{\mu=-1}^{1} \langle 1,m|\mu_x\rangle \langle\mu_x|\alpha\rangle \langle\ell| \hat D(-\mu\beta) |\nu_\alpha\rangle .
\label{eq:app_general_GRWA_coefficients}
\end{gather}

The displacement matrix elements in Eq.~\eqref{eq:app_general_GRWA_coefficients} are those collected in Appendix~\ref{app:analytical_details}. Once these coefficients are known, the quantities

\begin{gather}
    P_+ = \sum_\ell |\mathcal C_{1,\ell}|^2, \; P_0 = \sum_\ell |\mathcal C_{0,\ell}|^2, \; P_- = \sum_\ell |\mathcal C_{-1,\ell}|^2, \nonumber\\ Q = \sum_\ell \mathcal C_{1,\ell} \mathcal C_{-1,\ell}^{*}
\label{eq:app_PQ_general}
\end{gather}

determine the reduced density matrix, concurrence, and spin correlations through the formulas given in Sec.~\ref{subsec:reduced_observables}.

For an SGRWA eigenvector the approximation is identical, except that the physical oscillator amplitudes are obtained from

\begin{gather}
    \langle\ell| \hat D(-\mu\beta) \hat V_\lambda^\dagger |\nu_\alpha\rangle .
\label{eq:app_general_SGRWA_coefficients}
\end{gather}

Thus, all  observables are evaluated from the same retained eigenvector that determines the corresponding analytical energy and photon population. The fixed, self-consistent, and variational prescriptions differ only through the values of the bosonic parameters entering these expressions.

%%%%%%%%%%%%%%%%%%%%%%%%%%%%%%%%%%%%%%%%%%%%%%%%%%%%
\bibliography{references}
%%%%%%%%%%%%%%%%%%%%%%%%%%%%%%%%%%%%%%%%%%%%%%%%%%%%

\end{document}